\documentclass[useAMS,usenatbib]{mn2e}
\usepackage{graphicx}
\usepackage{paralist}

\newcommand{\aj}{\textnormal{AJ}}
\newcommand{\pasp}{\textnormal{PASP}}
\newcommand{\aap}{\textnormal{A\&A}}

\newcommand{\apj}{\textnormal{ApJ}}
\newcommand{\nat}{\textnormal{Nat}}
\newcommand{\mnras}{\textnormal{MNRAS}}
\newcommand{\apjs}{\textnormal{ApJS}}
\newcommand{\araa}{\textnormal{ARA\&A}}
\newcommand{\apjl}{\textnormal{ApJ}}

\title[The evolution of star formation activity in galaxy groups]{The evolution of star formation activity in galaxy groups}
\author[Erfanianfar dvi.]{G.Erfanianfar $^{1,2}$\thanks{E-mail:
erfanian@mpe.mpg.de}, P. Popesso$^{1,2}$, A. Finoguenov$^{3}$, S. Wuyts$^{2}$, D. Wilman$^{2}$,  A. Biviano$^{4}$, 
\newauthor F. Ziparo$^5$, M.~Salvato${^2}$, K.~Nandra$^{^2}$,  D.~Lutz$^{2}$, D.~Elbaz $^{6}$, M.~Dickinson$^{7}$, M. ~Tanaka$^{8}$,
\newauthor  M. Mirkazemi$^2$, M. L. Balogh$^{9,10}$, M B.~Altieri$^{11}$, H. Aussel$^{6}$, F. ~Bauer$^{12,13}$, S. Berta$^2$,
\newauthor  R. M. Bielby $^{14}$, N. Brandt$^{15}$, N. Cappelluti$^{16}$, A. Cimatti$^{17}$,M.~Cooper$^{18}$, D. Fadda$^{19}$, 
\newauthor   O. Ilbert$^{20}$, E. Le Floch$^{6}$, B. Magnelli$^{21}$, J. S. Mulchaey$^{22}$, R. Nordon$^{23}$, J. A. Newman$^{24}$, 
\newauthor  A. Poglitsch${^2}$, F. Pozzi$^{17}$\\
$^{1}$Excellence Cluster Universe, Boltzmannstr. 2, 85748 Garching bei M\"{u}nchen, Germany\\
$^{2}$Max-Planck-Institut f\"{u}r extraterrestrische Physik, Giessenbachstra\ss e 1, 85748 Garching bei M\"{u}nchen, Germany\\
$^{3}$University of Helsinki, Department of Physics, P.O. Box 64, FI-00014 Helsinki \\
$^{4}$INAF/Osservatorio Astronomico di Trieste, via G. B. Tiepolo 11, 34131 Trieste, Italy\\
$^{5}$School of Physics and Astronomy, University of Birmingham, Edgbaston, Birmingham B15 2TT, UK\\
$^{6}$Laboratoire AIM, CEA/DSM-CNRS-Universit{\'e} Paris Diderot, IRFU/Service d'Astrophysique,  B\^at.709, CEA-Saclay,\\
 91191 Gif-sur-Yvette Cedex, France.\\
$^{7}$National Optical Astronomy Observatory, 950 North Cherry Avenue, Tucson, AZ 85719, USA\\
$^{8}$National Astronomical Observatory of Japan, 2-21-1 Osawa, Mitaka, Tokyo 181-8588, Japan\\
$^{9}$Department of Physics and Astronomy, University of Waterloo, Waterloo, Ontario, N2L 3G1, Canada\\
$^{10}$Leiden Observatory, Leiden University, PO Box 9513, 2300 RA Leiden, The Netherlands\\
$^{11}$Herschel Science Centre, European Space Astronomy Centre, ESA, Villanueva de la Ca\~nada, 28691 Madrid, Spain\\
$^{12}$Instituto de Astrof́ısica, Facultad de F́ısica, Pontificia Universidad Catolica de Chile, 306, Santiago 22, Chile\\
$^{13}$Space Science Institute, 4750 Walnut Street, Suite 205, Boulder, CO 80301, USA\\
$^{14}$Dept. of Physics, Durham University, South Road, Durham, DH1 3LE, UK\\
$^{15}$Department of Astronomy and Astrophysics, 525 Davey Laboratory, The Pennsylvania State University, University Park, PA 16802, USA\\
$^{16}$INAF-Osservatorio Astronomico di Bologna, Via Ranzani 1, I-40127 Bologna, Italy\\
$^{17}$Dipartimento di Astronomia, Universit{\`a} di Bologna, Via Ranzani 1, 40127 Bologna, Italy\\
$^{18}$Center for Galaxy Evolution, Department of Physics and Astronomy, University of California, Irvine, 4129 Frederick\\
 Reines Hall Irvine, CA 92697, USA\\
$^{19}$NASA Herschel Science Center, Caltech 100-22, Pasadena, CA 91125, USA\\
$^{20}$LAM, CNRS-UNiv Aix-Marseille, 38 rue F. Joliot-Curis, F-13013 Marseille, France\\
$^{21}$Argelander-Institut f\"{u}r Astronomie, Universit\"{a}t Bonn, Auf dem H\"ugel 71, D-53121 Bonn, Germany\\
$^{22}$The Observatoires of the Carnegie Institution of Science, 813 Santa Barbara Street, Pasadena, CA 91101, USA\\
$^{23}$School of Physics and Astronomy, The Raymond and Beverly Sackler Faculty of Exact Sciences, Tel Aviv University, Tel Aviv 69978, Israel\\
$^{24}$Department of Physics and Astronomy, University of Pittsburgh and PITT-PACC, 3941 O'Hara St., Pittsburgh, PA 15260, USA\\
}

\begin{document}

\pagerange{\pageref{firstpage}--\pageref{lastpage}} \pubyear{2002}

\maketitle

\label{firstpage}
\clearpage
\begin{abstract}

We study the evolution of the total star formation (SF) activity, total stellar mass and halo occupation distribution in massive halos by using one of the largest X-ray selected sample of galaxy groups with secure spectroscopic identification in the major blank field surveys (ECDFS, CDFN, COSMOS, AEGIS). We provide an accurate measurement of SFR for the bulk of the star-forming galaxies using very deep mid-infrared Spitzer MIPS and far-infrared Herschel PACS observations. For undetected IR sources, we provide a well-calibrated SFR from SED fitting. We observe a clear evolution in the level of SF activity in galaxy groups. The total SF activity in the high redshift groups (0.5$<$z$<$1.1) is higher with respect to the low redshift (0.15$<$z$<$0.5) sample at any mass by $0.8\pm0.12$ dex. A milder difference ($0.35\pm0.1$ dex) is observed between the low redshift bin and the groups at $z \sim 0$.  We show that the level of SF activity is declining more rapidly in the more massive halos than in the more common lower mass halos. We do not observe any evolution in the halo occupation distribution and  total stellar mass- halo mass relations in groups. 
The picture emerging from our findings suggests that the galaxy population in the most massive systems is evolving faster than galaxies in lower mass  halos, consistently with a ``halo downsizing" scenario.

\end{abstract}

 \section{Introduction}

 One of the most fundamental correlations between the properties of galaxies in the local Universe is the so-called morphology-density relation (\citealt{Dressler1980,Davis1976}).
  A plethora of studies utilizing multi-wavelength tracers of activity have shown that late type star-forming galaxies favour low density regions in the local
 Universe (e.g. \citealt{Gomez2003, Blanton2005, Hogg2004, Lewis2002}). In particular, the cores of massive galaxy clusters are full of massive spheroids that are dominated by
 old stellar populations. A variety of physical processes might be effective in suppressing star formation and affecting the morphology of cluster and group
 galaxies. Two big families of such processes can be identified: (i) interactions with other cluster members and/or with the cluster potential and (ii)
 interactions with the hot gas that permeates massive galaxy systems. In the current standard paradigm for structure formation, dark matter collapses into halos
 in a bottom-up fashion: small objects form first and subsequently merge into progressively larger systems. In this context, galaxy groups are the €œbuilding
 blocks of galaxy clusters. Galaxy groups have at any epoch a volume density orders of magnitude higher than those of massive clusters, which represents the
 rare and extreme specimen at the high mass end of the dark halo mass function (\citealt{Jenkins2001}). This is confirmed by the observational evidence that groups
 are the most common environment of galaxies in the present day universe, containing 50\%-70\% of the galaxy population (\citealt{Geller1983, Eke2005}).
 This naturally implies that processes taking place in the group environment can have a significant impact on the evolution of the galaxy population as a whole.

The main debate now centers on the role of galaxy "internal" versus "external" processes as driving mechanisms of the galaxy evolution, or, according to an old-fashion
 approach, the ``nature" versus ``nurture" scenario. In the current paradigm of galaxy formation the ``internal" processes are mainly linked to the co-evolution of the
 host galaxy and its central black hole (\citealt{DiMatteo2005, Croton2006, Delucia2006, Hopkins2006}). However, as pointed out by \cite{Delucia2012}, the nature
 versus nurture dichotomy is an ill-posed problem. In the current paradigm of galaxy formation these physical internal and external
 processes are coupled with a history bias that is an integral part of the hierarchical structure formation of cosmic structure (\citealt{Delucia2012,Cooper2010}). \cite{Wilman2013} have
 demonstrated that halos in overdense regions statistically form earlier and merge more rapidly than halos in regions of lower density (\citealt{Gao2004}). This differential evolution leaves a trace on the observable
 properties of galaxies that inhabit different regions at any cosmic epoch (\citealt{Delucia2012}). This aspect makes the interpretation
 of the observational evidences even more difficult. In fact, binning galaxies according to their stellar mass does not suffice to disentangle the role of nature and nurture.
 For instance, two galaxies of identical mass at some cosmic epoch can end up having different stellar masses if one of them falls on to a cluster and the other remains
 in a region of average density. An important attempt to investigate from the observational point of view the inter-relationships between stellar mass, star-formation rate and
 environment comes from \cite{Peng2010} in the SDSS, zCOSMOS surveys. This study shows that a) two distinct processes, mass (internal) quenching and environment (external)
 quenching are both operating since z$\sim$1, b) environment-quenching occurs as large-scale structure develops and is more effective on satellite galaxies, c) mass-quenching
 is more efficient for central and generally more massive galaxies. The limit of this analysis is mainly in the definition of the environment that relies on the local galaxy
 density, which is only a poor proxy of the DM halo mass.

In the last decade a lot of effort has been devoted to the study of high redshift groups to investigate the possibility of a differential evolution of group galaxies with
 respect to field galaxies. A big step forward was made thanks to the advent of very deep multiwavelength surveys conducted on several blank fields, such as the Great
 Observatories Origin Deep Survey -South and -North (GOODS-S and GOODS-N, respectively), the Extended Chandra Deep Field South (ECDFS), the Cosmic Evolution Survey (COSMOS) and the
 All-wavelength Extended Groth strip International Survey (AEGIS). Those surveys combine deep photometric (from the X-rays to the far-infrared wavelengths) and spectroscopic
 (down to $i_{AB}$ $\sim$ 24 mag and b $\sim$ 25 mag) observations over relatively large areas to lead, for the first time, to the construction of statistically significant samples of groups up to
 to high redshift (z$\sim$1.3-1.6, e.g. \citealt{Finoguenov2010} and \citealt{Bielby2010}). In this context, the main outcome of these surveys is that group galaxies show a much faster
 evolution with respect to the field galaxies. For instance, the formation of the galaxy red sequence, which leads to the local dichotomy between red and blue galaxies, happens earlier in groups than in
 the field especially at high stellar masses (\citealt{Iovino2010, Kovach2010, Cooper2007, Wilman2009, Wilman2012}). It seems also that group galaxies undergo a substantial morphological transformation.
 Indeed, groups at z$\sim$1 host a transient population of "red spirals" which is not observed in the field (\citealt{Jeltema2007, Tran2008, Balogh2009, Wolf2009, Mei2012}). 

Most analyses so far have concentrated on comparisons of the star-forming properties of the group galaxy population as a whole with those of field galaxies. However,
 it is also important to assess the dependence (if any) of the star-forming properties of group galaxies on their system global properties, such as the mass, velocity dispersion
 and X-ray luminosity of the groups at different epochs to understand if and how the evolution of the star formation activity depends on these variables. A way of looking at the evolution of
 the SF activity in galaxy systems is to consider global quantities such as the total star formation rate, that is the sum of the SFRs of all the galaxies in a system (see e.g. \citealt{Popesso2007})
 or the fraction of star-forming galaxies in a system (see e.g. \citealt{Poggianti2006}). Understanding how the relation between these global quantities and the group properties
 changes with time can teach us how the evolution of galaxies depends on the environment where they live. For this purpose we create the largest homogeneously
 X-ray selected sample of groups at $0.15 < z < 1.1$ by using the deepest available X-ray surveys conducted with {$Chandra$} and {$XMM-Newton$} on the ECDFS, CDFN, COSMOS and EGS regions.
  In addition, we use the latest and deepest available {\it{Spitzer}} MIPS and {\it{Herschel}} PACS (Photoconducting Array Camera and Spectrometer, \citealt{poglitsch2010}) mid and
 far infrared surveys, respectively, conducted on the same blank fields to retrieve an accurate measure of the star formation rate of individual group galaxies. This is the first of a
 series of papers analyzing the relation between SF activity and galaxy environment defined as the membership of a galaxy to a massive dark matter halo. In this paper we carefully
 describe the catalog and present a calibration of all the relevant quantities involved in our analysis. We use this unprecedented dataset to study the evolution of the relation between
 the total SFR in galaxy groups at $0.15 < z < 1.1$ with the group global properties, mainly the total halo mass, and to the stellar mass content of the groups and Halo Occupation Distribution
 (HOD) to understand how the group galaxy population evolves though cosmic times.

The paper is organized as follows. In Sect. 2 we describe our dataset. In Sect. 3 we describe how all relevant quantities are estimated.
 In Sect. 4 we describe our results and in Sect. 5 we discuss them and draw our conclusions. We adopt a \cite{Chabrier2003}
 initial mass function (IMF), H$_0=71$ km~s$^{-1}$~Mpc$^{-1}$, $\Omega_m=0.3$, $\Omega_{\Lambda}=0.7$ throughout this paper.

\section[]{The dataset}

The aim of this work is to analyse the evolution of the star formation activity in galaxy groups.
 For this purpose, we build a dataset which combines wide area surveys with good X-ray coverage, deep photometry,
 and high spectroscopic coverage. Galaxy clusters and groups are permeated by a hot intracluster medium
 radiating optically diffuse thermal emission in the X-ray band. Under the condition of hydrostatic equilibrium,
 the gas temperature and density are directly related to halo mass. A tight relation (rms$\sim$0.15 dex) exists also between the cluster
 dynamical mass and the X-ray luminosity (L$_X$, \citealt{Pratt2007, Rykoff2008}). A similar scaling relation, though with a larger scatter,
 holds also in the galaxy group mass region (\citealt{Sun2012}, rms$\sim$0.3 dex). Thus, the X-ray  selection is the best way
 to select galaxy groups and clusters and to avoid incorrect galaxy group identifications due to projection effects associated with optical selection techniques.
 In addition, deep and accurate multi-wavelength catalogues are necessary in order to identify the group membership and to study the properties of the group galaxy population.
  Thus, we combine X-ray selected group catalogues and photometric and spectroscopic galaxy catalogues of four major blank field surveys: AEGIS, COSMOS, ECDFS and CDFN.
 Throughout our analysis we use spectroscopic redshifts to define the group membership and the multiwavelength photometric information for studying the galaxy properties.
 For calibration purposes we will also make use of photometric redshifts. In the following section we describe the multiwavelength dataset of each field.

\subsection{The blank fields}

\subsubsection{AEGIS}
The All-Wavelength Extended Groth Strip International Survey (AEGIS) brings together deep imaging data from X-ray to
 radio wavelengths and optical spectroscopy over a large area (0.5-1 $deg^2$; \citealt{Davis2007}). This survey includes: $Chandra$/ACIS X-ray
 (0.5-10 keV; \citealt{Laird2009}), GALEX ultraviolet (1200-2500 $\AA$), CFHT/MegaCam Legacy Survey optical (3600-9000 $\AA$), CFHT/CFH12K optical
 (4500-9000 $\AA$; \citealt{Coil2004}), Hubble Space Telescope/ACS optical (4400-8500 $\AA$; \citealt{Lotz2008}), Palomar/WIRC near-infrared (1.2-2.2 $\mu m$;\citealt{Bundy2006}),
 Spitzer/IRAC mid-infrared (3.6, 4.5, 5.8, 8 $\mu m$; \citealt{Barmby2008}), Herschel far-infrared ( 100, 160 $\mu m$), VLA radio continuum (6-20cm; \citealt{Willner2012}) and a large spectroscopic dataset. 

In particular, the X-ray data come from sensitive $Chandra$ and $XMM-Newton$ observations of this field which lead to one of the largest X-ray selected samples of galaxy groups catalog to date \citep{Erfanianfar2013}.
 The total X-ray exposure time with $Chandra$ in this field is about 3.4 Ms with a nominal exposure of 800 ks in three central
 fields. The $XMM-Newton$ observations in the southern part of this field have an exposure of 100 ks. 
The spectroscopic information is taken from different spectroscopic surveys performed in this field. The AEGIS field,
 as part of the Extended Groth Strip (EGS) field, has been targeted with the DEEP2 galaxy redshift survey (\citealt{Davis2003,Newman2012})
 and it is the only field that has been a subject of extensive spectroscopic follow-up data in DEEP3 (\citealt{Cooper2011,Cooper2012}).
 In the DEEP2 fields EGS is the only field which is not color selected, so it gives us a nearly complete sample with redshift.
 In addition to DEEP2 and DEEP3, EGS is located in Sloan Digital Sky Survey coverage so we have additional spectra for low-redshift galaxies.
 We also used redshifts of spectroscopic galaxies obtained in follow-up observations of 
the DEEP2 sample with the Hectospec spectrograph on the Multiple Mirror Telescope (MMT; \citealt{Coil2009}).

Furthermore, the EGS field is located at the center of the third wide field of the Canada-France-Hawaii Telescope Legacy Survey (CFHTLS-Wide3, W3)
 which is imaged in $u^\ast$, $g^\prime$, $r^\prime$, $i^\prime$ and $z^\prime$ filters down to $i^\prime$=24.5 with photometric data for 366,190
 galaxies (\citealt{Brimioulle2008}). The EGS field also contains the CFHTLS Deep 3 field, which covers 1 deg$^2$ with $ugriz$ imaging
 to depths ranging from 25.0 in $z$ to 27 in $g$. For this work, we have used the T0006 release \footnote{http://terapix.iap.fr/cplt/T0006-doc.pdf} of the CHTLS Deep data. The CFHTLS Deep field also contains near-infrared
 coverage in the $JHK$ bands via the WIRCam Deep Survey (WIRDS - \citealt{Bielby2012}). This covers 0.4 deg$^2$ of the D3 field and provides deep imaging to $\sim24.5$ (AB)
 in the three NIR bands. Photometric redshifts in the region covered by the NIR data were determined using the \texttt{Le Phare} code as described in \cite{Bielby2012}.
 
\subsubsection{COSMOS}

The Cosmological Evolution survey (COSMOS) is the largest survey ever made using the Hubble Space Telescope (HST).
 With its 2 square degrees of coverage, COSMOS enables the sampling of the large scale structure of the universe,
 and reduces cosmic variance (\citealt{Scoville2007}). In particular COSMOS guarantees full spectral coverage, with X-ray ($Chandra$ \& $XMM-Newton$),
 UV (GALEX), optical (SUBARU), near-infrared (CFHT), mid-infrared (Spitzer), Herschel far-infrared ( 100, 160 $\mu m$), sub-millimetric (MAMBO) and radio (VLA) imaging. Furthermore,
the X$-$ray information provided by the 1.5 Ms exposure with $XMM-Newton$
(53 pointings on the whole field, 50 ksec each, \citealt{Hasinger2007}) and the additional 1.8 Ms exposure with $Chandra$
 in the central square degree \citep{Elvis2009} enable robust detections of galaxy groups out to $z\sim 1.2$ \citep{Finoguenov2007, George2011, George2013}.

 COSMOS has been targeted by many spectroscopic programs at different telescopes and has a broad spectral coverage. 
The spectroscopic follow up is still continuing and so far includes: the zCOSMOS survey at VLT/VIMOS \citep{Lilly2007,Lilly2009}, GEEC2 survey with the GMOS spectrograph
 on the GEMINI telescope \citep{Balogh2011,Mok2013}, Magellan/IMACS \citep{Trump2007} and MMT \citep{Prescott2006}
 campaigns, observations at Keck/DEIMOS (PIs: Scoville, Capak, Salvato, Sanders, Kartaltepe) and FLWO/FAST \citep[][]{Wright2010}.

The COSMOS photometric catalog \citep{Capak2007,Capak2009} contains multi-wavelength photometric information for $\sim 2\times 10^6$
 galaxies over the entire field. The position of galaxies has been extracted from the deep i-band imaging \citep{Taniguchi2007}.
  A limit of 80\% completeness is achieved at $i_{\rm AB}=$26.5. The optical catalog of  \cite{Capak2007,Capak2009} includes
 31 bands (2 bands from the Galaxy Evolution Explorer (GALEX), 6 broad bands from the SuprimeCam/Subaru camera, 2 broad bands
 from MEGACAM at CFHT, 14 medium and narrow bands from SuprimeCam/Subaru, J band from the WFCAM/UKIRT camera, H and K band from
 the WIRCAM/CFHT camera, and the 4 IRAC/Spitzer channels). In particular, We take the catalogue provided by \cite{Ilbert2009}
 and \cite{Ilbert2010}. They cross-match the S-COSMOS \cite{Sanders2007} 3.6 $\mu$m selected catalogue with the multi-wavelength
 catalogue \citep{Capak2007,Capak2009} and calculate photo-z, stellar masses and SFR in a consistent way by using the \texttt{Le Phare} code\citep{Ilbert2009, Ilbert2010}.

\subsubsection{ECDFS}
  ECDFS is observed broadly from  X-ray to radio wavelengths and centred on one of the most well-studied extragalactic fields in the sky (e.g. \citealt{Giavalisco2004}; \citealt{Rix2004};
 \citealt{Lehmer2005}; \citealt{Quadri2007}; \citealt{Miller2008}; \citealt{Padovani2009}; \citealt{Cardamone2010}; \citealt{Xue2011};
 \citealt{Damen2011}). The smaller Chandra Deep Field South
 (CDFS, $\alpha = 03h 32m25s$ , $\delta = -27^o 49 ^m 58^s$), in the central part of ECDFS, is currently the deepest X-ray survey with $Chandra$ (4Ms) and $XMM-Newton$ (3Ms) programs. 

The redshift assemblage in the ECDFS and the smaller CDFS and GOODS-S regions is achieved by complementing the spectroscopic redshifts
 contained in the \cite{Cardamone2010} catalog with all new publicly available spectroscopic redshifts, such as the one of
 \cite{Silverman2010} and the Arizona CDFS Environment Survey (ACES, \citealt{Cooper2012}). We clean the new catalogue of redshift
 duplications for the same source by matching the \cite{Cardamone2010} catalog with the \cite{Cooper2012} and the \cite{Silverman2010} catalog
 within $1\arcsec$ and by keeping the most accurate ${\rm z_{spec}}$ entry (smaller error and/or higher quality flag) in case of multiple
 entries (see \citealt{Ziparo2013} for a more detailed discussion). We also include the very high quality redshifts
 of the GMASS survey (\citealt{Cimatti2008}) using the same procedure. The total number of secure redshifts in the sample is 5080 out of 7277 total, unique targets.

We use the multi-wavelength photometric data from the catalogue of  \cite{Cardamone2010}. It includes a total of 10 ground-based 
broad bands ($U$, $U38$, $B$, $V$, $R$, $I$, $z$, $J$, $H$, $K$), 4 IRAC bands ($3.6~\mu$m, $4.5~\mu$m, $5.8~\mu$m, $8.0~\mu$m),
 and 18 medium-band imaging ($IA427$, $IA445$, $IA464$, $IA484$, $IA505$, $IA527$, $IA550$, $IA574$, $IA598$, $IA624$, $IA651$, $IA679$, $IA709$, $IA738$, $IA767$, $IA797$, $IA856$).
 The catalogue includes multi-wavelength SEDs and photometric redshifts for $\sim 80000$ galaxies down to $\rm{R_{AB}} \sim 27$.

\subsubsection{CDFN}
The Chandra Deep Field North (CDFN) survey is one of the deepest 0.5-8.0 keV surveys ever made. The Chandra survey is comprised of two partially overlapping $\sim$1 Ms ACIS-I exposures covering a total of 448 sq. arcmin, of which $\approx$160 sq. arcmin has 1.7-1.9 Ms of exposure. In addition, there is 150 ks of good XMM-Newton exposure.
 The GOODS-North field within the CDFN centers at RA$=12^{\rm h} 36^{\rm m} 55^{\rm s}, $
Dec.$=+62^\circ 14^{\rm m} 15^{\rm s} $ (J2000) and has become one of the most well-studied extragalactic fields in the sky with existing
 observations among the deepest at a broad range of wavelengths (e.g., \citealt{alexander2003}; \citealt{morrison2010}; \citealt{Cooper2011}; \citealt{Elbaz2011}).
 GOODS-N covers an area of approximately $10\arcmin \times 16 \arcmin$ \citep{Giavalisco2004}.
 
We use the multi-wavelength catalogue of GOODS-N built by \cite{Berta2010}, who adopted the \cite{Grazian2006} approach for the PSF matching. The catalogue includes ACS $bviz$ (\citealt{Giavalisco2004}),
 Flamingos $JHK$, and Spitzer IRAC data. Moreover, MIPS 24 $\mu$m (\citealt{Magnelli2009}) and deep $U$, $Ks$ (\citealt{Barger2008}).
 The catalog is also complemented by the spectroscopic redshift compilation of \cite{Barger2008}.

\subsection{X-ray Analysis}
\label{Xray}

All the blank fields considered in our analysis are observed extensively in the X-ray with $Chandra$ and $XMM-Newton$.
Firstly, we remove point sources in both of the $Chandra$ and $XMM-Newton$ images following the procedure explained in \cite{Finoguenov2009}. Then the residual images were coadded,
 taking into account the different sensitivity of each instrument.
  The ``residual" image, free of point sources, is then used to identify extended emission. Groups and clusters are selected as extended emission with at least 4$\sigma$
 significance with respect to the background (see \citealt{Finoguenov2009} for further details regarding the precise definition of background and detection significance level). A redshift to each system on the basis of spectroscopic redshift, when available, or otherwise photometric redshift is assigned. The X-ray luminosity  $\mathrm{L_X}$  and $\mathrm{r_{500}}$ is determined iteratively, based on the aperture flux and recalculating the correction for the missing flux.
 M$_{200}$ is determined via the scaling relation from weak lensing by the final L$_X$ and so is r$_{200}$. The r$_{200}$ is the radius at which the density of a cluster is equal to 200 times the critical density of the universe ($\mathrm{\rho_c}$) and is defined as $\mathrm{M_{200}=(4 \pi/3) 200 \rho_c r_{200}^3}$. After taking into account the possible missed flux through the use of the beta-model. The total masses $\mathrm{M_{200}}$,
 within $\mathrm{r_{200}}$, are estimated based on the measured $\mathrm{L_X}$ and its errors, using 
the scaling relation from weak lensing calibration of \cite{Leauthaud2010}. The intrinsic scatter for mass in this relation is 20\% \citep{Leauthaud2010, Allevato2012} which is larger than a formal statistical error associated
 with the measurement of $\mathrm{L_X}$.

The X-ray group catalogs derived with this approach comprise 52 detections in AEGIS (\citealt{Erfanianfar2013}), 277 detections in the COSMOS field (\citealt{George2011}),
 50 detections in the ECDFS (Finoguenov et al. in prep.) and 27 detection in CDFN. We present the full CDFN X-ray group catalog in the Appendix.
 In the following section we describe how we select a subsample of ``secure'' groups and how we associate them to the respective galaxy population.

\subsubsection{Group Identification}

To associate the respective galaxy population to any X-ray extended emission and to define properly the group redshift we follow the same procedure described in \cite{Erfanianfar2013}
 and performed on the AEGIS X-ray dataset. We extend here this procedure to all the other fields described in the previous section.
 In brief, we estimate the galaxy over-density along the line of sight in the region of each X-ray extended emission following the red sequence technique (\citealt{Finn2010}).
 Additionally we screen for the existence of an over-density of red galaxies in the 3rd dimension using the spectroscopic redshift distribution of the X-ray extended source.

As described in \cite{Erfanianfar2013}, we assigned to each X-ray extended source a flag that describes the quality of the identification. We define the following flags:
\begin{itemize}
\item [-] Flag=1 indicates a confident redshift assignment, significant X-ray emission, and a well-determined center of red galaxies with respect to X-ray emission center
\item [-] Flag=2 indicates that the centering has a large uncertainty ($\sim$ $15''$)
\item [-] Flag=3 indicates no secure spectroscopic confirmation but good centering
\item [-] Flag=4 or more depending on the survey indicates that we have uncertain redshifts due to the lack of spectroscopic objects and red galaxies,
 and also a large uncertainty in centering or unreliable cases for which we could not identify any redshift. 

\end{itemize}


For the purpose of this work we consider only X-ray extended emission with a secure redshift definition with flag 1 or 2. Out of the initial
 406 X-ray group candidates in the four considered fields, we identify 244 secure groups. The secure redshift estimate is used to refine the initial
 X-ray luminosity of the groups and, thus, the mass M$_{200}$ with the scaling relation of \cite{Leauthaud2010} as described in the previous paragraph.
 The final step of the analysis is the identification of the group galaxy members via dynamical analysis as described below.

 \subsubsection{Group Membership}
\label{membership}

In order to properly define the galaxy membership of each group, we identify among our 244 secure groups those which are relatively isolated.
 Indeed, the presence of a close companion may bias the estimate of the velocity dispersion of the group and, thus, also the galaxy membership
 definition which relies on this quantity. This procedure leads to a subsample of 211 clean isolated groups. 
We follow the procedure described in \cite{Erfanianfar2013} to estimate the group velocity dispersion and the galaxy membership definition.
 The procedure is iterative and it needs a first guess of the velocity dispersion to define the redshift interval around the group redshift to determine the initial galaxy membership.
 We derive the first guess of the velocity dispersion from the group's X-ray luminosity $L_X$ by using the relation of \citealt{Leauthaud2010}. This velocity dispersion provides
 the intrinsic velocity dispersion ($\sigma(v)_{intr}$-which can be achieved by subtracting the errors of the redshift measurements in
quadrature from the rest frame velocity dispersion) of the group. We estimate, then, the observed velocity dispersion by considering the redshift of the group ($z_{group}$) and the
 errors of the redshift in our spectroscopic samples, $\langle\Delta(v)\rangle^2$ according to these relations:
 
 \begin{equation}
 \label{2}
 \sigma(v)_{rest}^2=\sigma(v)_{intr}^2+\langle\Delta(v)\rangle^2
 \end{equation}
 
 \begin{equation}
 \label{1}
 \sigma(v)_{obs}=\sigma(v)_{rest}\times(1+z_{group})
 \end{equation}
 
\begin{figure}
 \includegraphics[width=80mm]{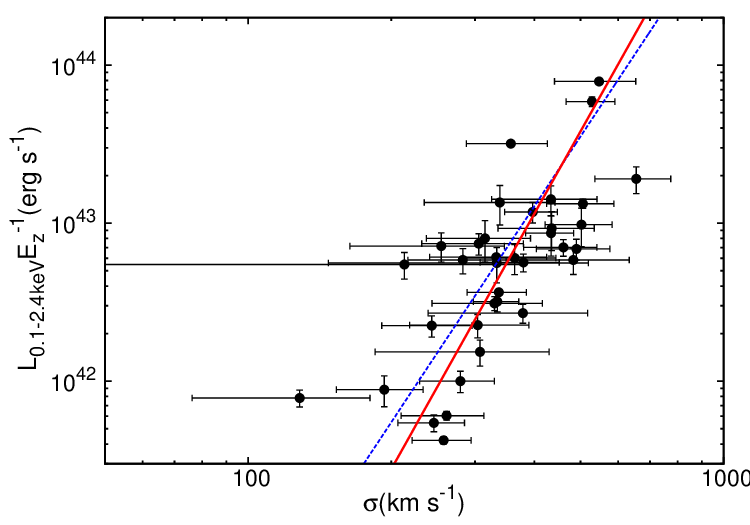} 
  \caption{$L_X-\sigma$ relation for X-ray groups. The dashed blue line show our expectation for $L_X-\sigma$ relation from scaling relations (\citealt{Leauthaud2010})
 and the solid red line is our bisector fit to data.}
\label{fig1}
 \end{figure}

We consider as initial group members all galaxies within $|z-z_{group}|<\delta(z)_{max}$ where $\delta(z)_{max}=2\frac{\sigma(v)_{obs}}{c}$ and within virial radii ($r_{200}$) from the X-ray center.
 We recompute the observed velocity dispersion of the groups, $\sigma(v)_{obs}$ using the ``gapper'' estimator method which gives more accurate measurements of velocity
 dispersion for small size groups (\citealt{Beers1990,Wilman2005a}) in  comparison to the usual formula for standard deviation (see \citealt{Erfanianfar2013} for more details).
 The observed velocity dispersion is estimated according to:
 \begin{equation}
 \label{eqn:optfilt}
 \sigma(v)_{obs}=1.135c\times\frac{\sqrt\pi}{N(N-1)}\sum_{i=1}^{n-1}\omega_ig_i
 \end{equation}
 where $w_i=i(N-i)$ , $g_i=z_{i+1}-z_i$ and $N$ is the total number of spectroscopic members. In this way we measure the velocity dispersion using the line-of-sight velocity gaps where
 the velocities have been sorted into ascending order. The factor 1.135 corrects for the 2$\sigma$ clipping of the Gaussian velocity distribution. We iterate the entire
 process until we obtain a stable membership solution. We then calculate errors for our velocity
dispersions using the Jackknife technique (\citealt{Efron1982}). The procedure can be considered reliable for groups with at least 10 galaxy members.
 The 10 galaxy members threshold is reached for 36 groups out of 211. For the groups with less than 10 members but still more than 5 members within $r_{200}$, we base the velocity dispersion estimate on L$_X$ and the
 relation between $\sigma$ and L$_X$ as in \citealt{Leauthaud2010}. This leads to a sample of 111 groups out of 211.
 Figure \ref{fig1} shows the $L_x-\sigma$ relation for X-ray groups with more than 10 spectroscopic members, where $\sigma$ is estimated via dynamical analysis.
 The solid red line shows the power-law fit to the relation. The bisector procedure is used for this fit (\citealt{Akritas1996}). We also plot the $L_x-\sigma$ relation (dashed blue line) expected
 from scaling relations obtained for a sample of groups with similar luminosities in the $0<z<1$ redshift range in COSMOS (\citealt{Leauthaud2010}). 
The consistency between two relations ensures that the estimate of the velocity dispersion derived from the X-ray luminosity and the one calculated via dynamical analysis are in good agreement.

Once we have the estimate of the velocity dispersion of each group, we define as group members all galaxies within $2\times r_{200}$  in the angular direction and
 $\pm3\times(\sigma/c)\times(1+z_{group})$ in the line of sight direction in order to consider also the infalling regions of the groups. When a member galaxy is associated
 to more than one group, we consider it as a member of the group for which the distance to the galaxy is lowest in units of virial radii.

 \subsection{Infrared data}

For all considered fields we use the deepest available {\it{Spitzer}} MIPS 24 $\mu$m and PACS 100 and 160 $\mu$m datasets. For COSMOS, these are coming from
 the public Spitzer 24 $\mu$m \citep{LeFloch2009,Sanders2007} and PEP PACS 100 and 160 $\mu$m data (\citealt{Lutz2011}).  Both  {\it{Spitzer}} MIPS 24 $\mu$m and PEP
 source catalogues are obtained by extracting sources using NIR priors as described in \cite{Magnelli2009}. In short, IRAC and MIPS 24~$\mu$m source positions are used
 to detect and extract MIPS and PACS sources, respectively, at 24, 100 and 160~$\mu$m. This is feasible since extremely deep IRAC and MIPS 24~$\mu$m observations
 are available for the COSMOS field (\citealt{Scoville2007}). The source extraction is based on a PSF-fitting technique, presented in detail in \cite{Magnelli2009}.
 The association between 24 $\mu$m and PACS sources with their optical counterparts, taken from the optical catalog of \cite{Capak2007} is done via a maximum
 likelihood method \citep[see][for details]{Lutz2011}. 

The same approach is used also for the AEGIS field, where we use the {\it{Spitzer}} MIPS 24 and PEP PACS 100 and 160 $\mu$m catalogs produced by the PEP team (see \citealt{Magnelli2009}).

In the CDFS and GOODS regions the deepest available MIR and FIR data are provided by the  {\it{Spitzer}} MIPS 24 $\mu$m Fidel Program  (\citealt{Magnelli2009}) and by the
 combination of the PACS PEP (\citealt{Lutz2011}) and GOODS-Herschel (\citealt{Elbaz2011}) surveys at 70, 100 and 160 $\mu$m. The GOODS Herschel survey covers a smaller
 central portion of the entire GOODS-S and GOODS-N regions. Recently the PEP and the GOODS-H teams combined the two sets of PACS observations to obtain the deepest
 ever available PACS maps (\citealt{Magnelli2013}) of both fields. The more extended CDFS area has been observed in the PEP survey as well, yet having a higher flux limit.
 As for the COSMOS catalogs, the 24 $\mu$m and PACS sources are associated to their optical counterparts via a maximum likelihood method \citep[see][for details]{Lutz2011}. 

For all galaxies identified as galaxy group members, we use the MIPS and PACS data to accurately estimate the IR bolometric luminosity and, thus, the SFR. We compute
 the IR luminosities integrating the SED templates from \cite{Elbaz2011} in the range 8-1000\,$\mu$m. The PACS (70, 100 and 160\,$\mu$m) fluxes, when available, together
 with the 24 $\mu$m fluxes are used to find the best fit templates among the main sequence (MS) and starburst (SB, \citealt{Elbaz2011})  templates. When only the 24 $\mu$m
 flux is available for undetected PACS sources, we rely only on this single point and we use the MS template for extrapolating the $L_{IR}$. Indeed, \cite{Ziparo2013}
 show that the MS template turns out to be the best fit template in the majority of the cases with common PACS and 24 $\mu$m detection. \cite{Ziparo2013} show also that
 by using only 24 $\mu$m data and the MS template there could be a slight underestimation (10\%) only above $z\sim 1.7$ or $L_{IR}^{24} > 10^{11.7}\ L_\odot$. In larger
 fields such as COSMOS and ECDFS there is a larger probability to find rare strong star-forming off-sequence galaxies at $L_{IR}^{24} > 10^{11.7}\ L_\odot$  even at low
 redshift. However, those sources should be captured by the Herschel observations given the very high luminosity threshold. Thus, it would not be a problem in getting
 a proper estimate of the $L_{IR}$ from the best fit templates also for these rare cases. The SFR for these sources is then estimated via the \cite{Kennicutt1998} relation
and then corrected from Salpeter IMF to Chabrier IMF for consistency with SFR$_{SED}$ and stellar mass.

 \begin{figure*}
 \includegraphics[width=130mm]{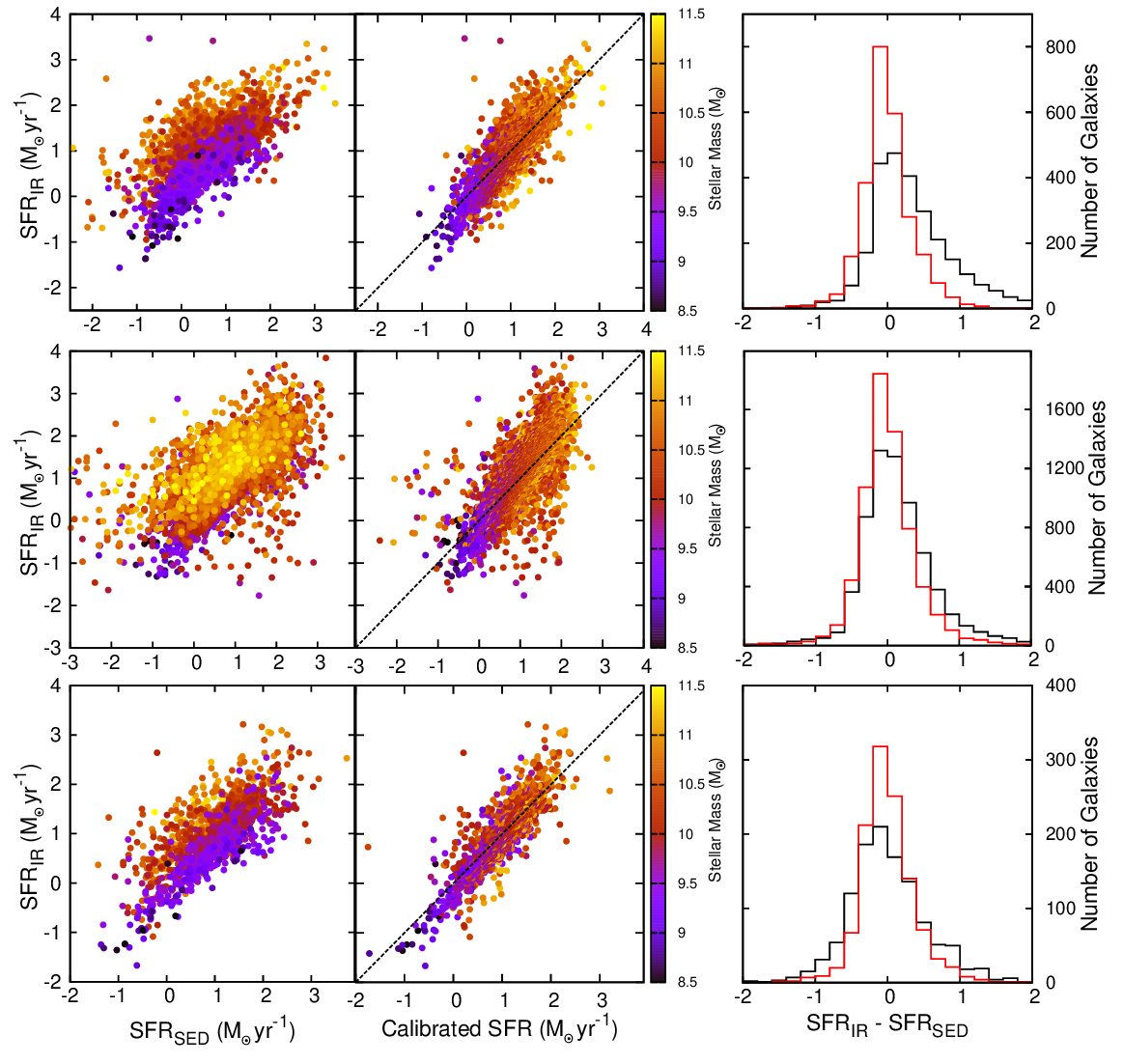}
  \caption{The left panels show SFR$_{IR}$ vs. SFR$_{SED}$ color-coded by stellar mass before re-calibration for EGS, COSMOS and GOODS-S from top to bottom respectively. The middle panels show corresponding
SFR$_{IR}$ vs.re-calibrated SFR$_{SED}$. The dashed line is one to one relation. The right panels are the histogram of corresponding SFR$_{IR}-SFR_{SED}$. The black and red histograms
 show before (black) and after (red) re-calibration.}
\label{fig2}
 \end{figure*}

\begin{figure*}
\centering
 \includegraphics[width=0.4\textwidth]{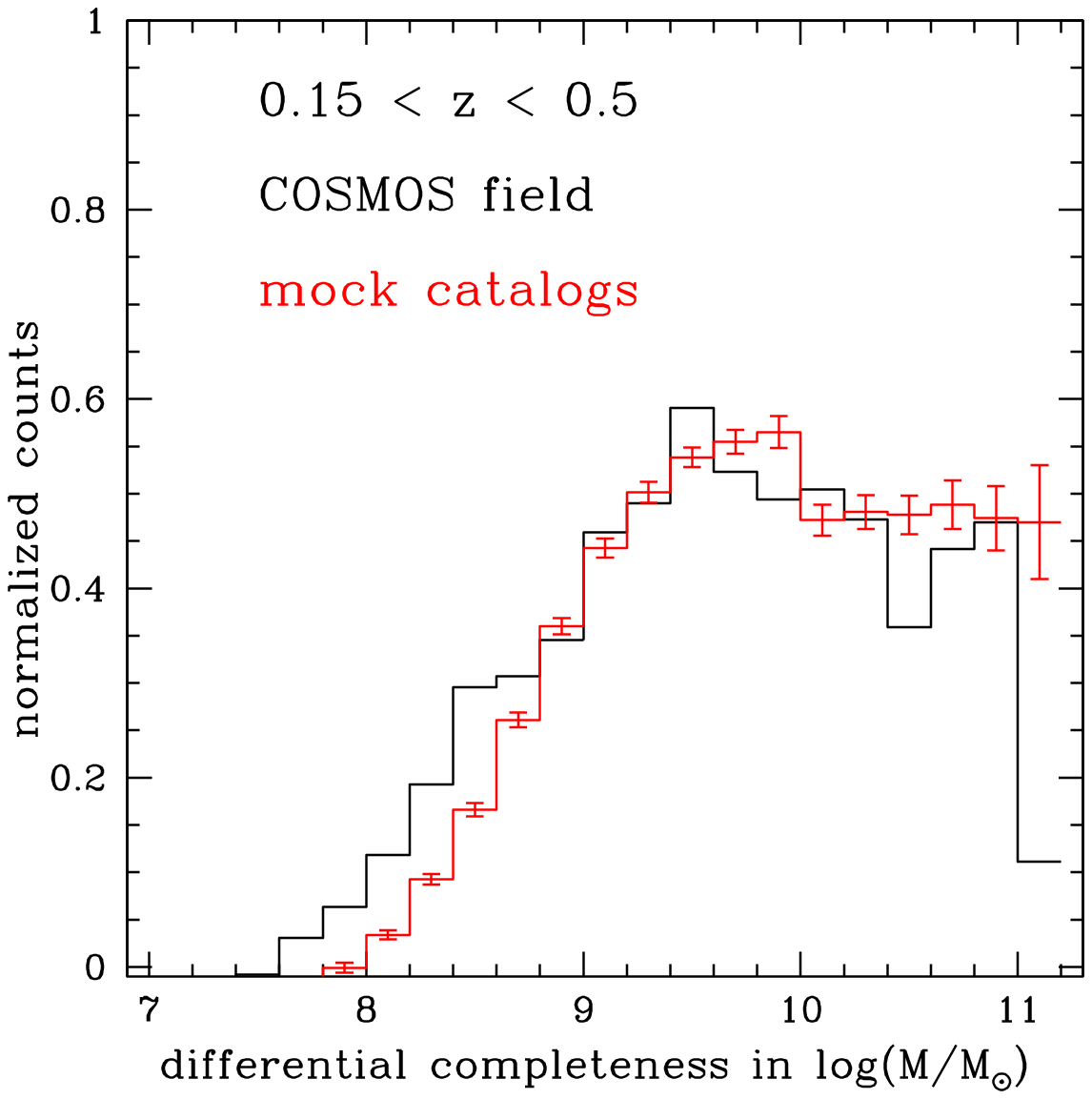}
 \includegraphics[width=0.4\textwidth]{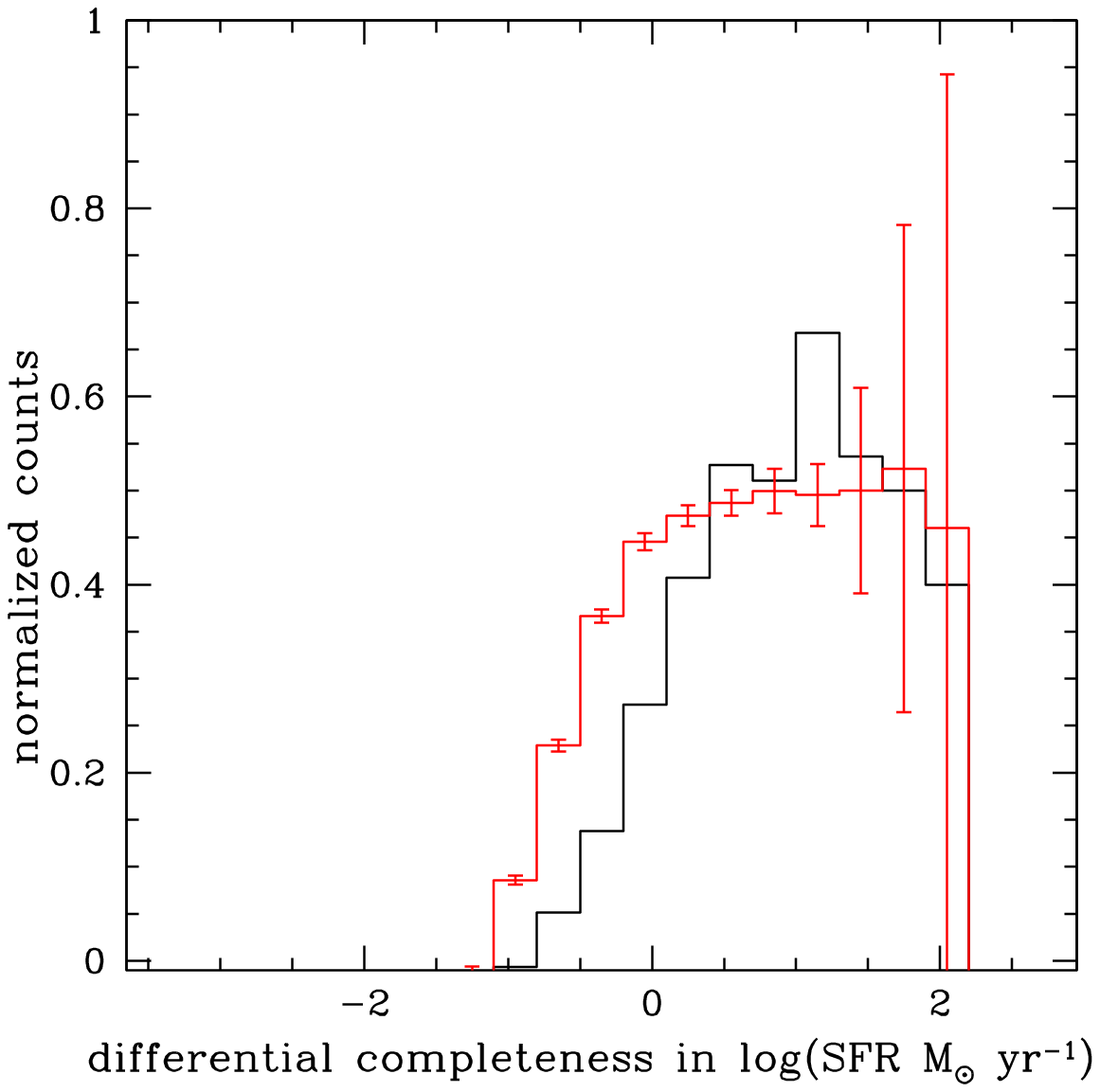}
  \caption{{\it{Left panel}}: spectroscopic completeness per stellar mass bin in the low redshift sample (black histogram) and
 in the simulated ``incomplete'' mock catalog (red line) in the same redshift range. {\it{Right panel}}: spectroscopic completeness
 per SFR bin for the low redshift sample (black histogram) and in the simulated ``incomplete'' mock catalog (red line).}
\label{inco}
 \end{figure*}

\subsubsection{Stellar masses and star formation rate from SED fitting}
 \label{sed_fitting}

Due to the flux limits of the MIPS and PACS catalogs in the four considered blank fields,
 the IR catalogs are sampling only the Main Sequence region and can not provide a SFR estimate
 for galaxies below the Main Sequence or in the region of quiescence. For a complete census of
 the star formation activity of the group galaxies, we need, however, an estimate of the SFR
 of all group members. For this reason, we complement the SFR estimates derived from IR data
 (SFR$_{IR}$), as described in the previous section, with an alternative estimate of the SFR.
SFR based either on SED fitting technique (SFR$_{SED}$) or on rest-frame UV observations (SFR$_{UV}$) are both reliable candidates. According to \cite{Ziparo2013}, the scatter
 of the SFR$_{UV}$-SFR$_{IR}$ relation is always bigger (at every redshift) with respect to the SFR$_{SED}$-SFR$_{IR}$ calibration. So, we use SFR$_{SED}$ as an 
alternative estimate of the SFR. Thus, for all galaxies undetected
 in MIPS and PACS maps, we use the SFR$_{SED}$ taken
from the following catalogs:
\begin{itemize}
\item[-] in AEGIS, SFR estimated with FAST \citep{Kriek2009} taken from \cite{Wuyts2011}
\item [-] in COSMOS, SFR estimated with \texttt{Le Phare} taken from \cite{Ilbert2010}
\item [-] in ECDFS, SFR estimated with \texttt{Le Phare}, from \cite{Ziparo2013}
\item [-] in CDFN, SFR estimated with FAST \citep{Kriek2009} taken from \cite{Wuyts2011}
\end{itemize}

The same catalogs provide also an estimate of the galaxy stellar mass. All SFR$_{SED}$ and stellar mass estimates are in Chabrier IMF.

\cite{Ziparo2013} point out, in general, the stellar masses and SFR$_{SED}$ derived from \cite{Wuyts2011}, \cite{Ilbert2010} and \cite{Ziparo2013}
 are all in agreement when compared on a common galaxy subsample.According to \cite{Ziparo2013}, the scatter around the 1 to 1 relation is of the order of 0.6 dex.
 Indeed, previous studies \citep{Papovich2001,Shapley2001,Shapley2005,Santini2009} already demonstrate that,
 while stellar masses are rather well determined (within a factor of 2) by very different methods, the SED fitting procedure does not strongly constrain star
 formation histories at high redshifts, where the uncertainties become larger due to the SFR--age--metallicity degeneracies. This degeneracy leads to the confusion
 of young, obscured star-forming galaxies with more massive, old, more quiescent galaxies. \cite{Wuyts2011} confirm the SFR$_{SED}$ provides a quite good
 estimate of the SFR for moderately star-forming galaxies and fails to provide a good estimate for very obscured objects.

Indeed, if we examine the scatter of the SFR$_{SED}$$-$SFR$_{IR}$ relation we clearly see a degeneracy with the stellar mass, as shown in the left column panels of Figure \ref{fig2}.
 This degeneracy is stronger than the one due to the redshift, as shown in \cite{Wuyts2011}, though the two aspects are related via selection effects
 (only massive star-forming galaxies are generally have spectroscopic redshifts at high redshift). The mass dependence of the scatter is different from field
 to field and depends on the method used for the SED fitting. This is probably due to two aspects. First, any blank field is characterized by a different
 dataset in terms of multiwavelength  coverage (number and type of broad band filters) and, thus, by a different sampling of the galaxy SED. Second, different
 recipes, thus different star formation histories,  and different fitting techniques are used for estimating the stellar masses and the SFR$_{SED}$.
  This also explains why there is such a large scatter in the SFR$_{SED}$ derived with different methods.

The result of this exercise shows that we can not use the  SFR$_{SED}$$-$SFR$_{IR}$ relation observed in one of the fields to calibrate the SFR$_{SED}$ of the
 other fields or obtained with a different method. Thus, we use the following approach. In order to correct a posteriori for the stellar mass bias in the
 SFR$_{SED}$, we fit the plane SFR$_{IR}-$SFR$_{SED}-$Mass, separately for each field. The best fit relation is listed below for AEGIS and CDFN (same fitting procedure):
\begin{equation}
 SFR_{IR}=-6.16+0.59 \times SFR_{SED}+0.66 \times M_{*}
\end{equation}
for COSMOS:
\begin{equation}
 SFR_{IR}=-4.54+0.61 \times SFR_{SED}+0.49 \times M_{*}
\end{equation}
and for ECDFS and GOODS-S:
\begin{equation}
 SFR_{IR}=-4.56+0.63 \times SFR_{SED}+0.49 \times M_{*}
\end{equation}

Once this calibration is used to correct the SFR$_{SED}$ with the additional information of
the stellar mass, the scatter around the SFR$_{SED}-$SFR$_{IR}$ relation decreases to 0.21 dex, 0.23 dex
 and 0.12 dex in comparison to SFR$_{IR}$ for galaxies with more than $10^{10} M_{\odot}$ in AEGIS, COSMOS,
 and ECDFS, respectively, as shown in central column panels of Figure \ref{fig2}. The values of the scatter
 are still 0.34, 0.42 and 0.44 in AEGIS, COSMOS, and ECDFS, respectively, when the whole mass range is considered.

We adopt this calibration to correct a posteriori the SFR$_{SED}$ estimates for all IR undetected galaxies above $log(SFR)>-0.5$.
 We think that this calibration is applicable in the SFR range considered here to IR undetected galaxies for the following reasons. \cite{Elbaz2011}
 show that the IR SED of star-forming galaxies are not evolving with redshift and that, instead, there is a much stronger dependence on the location
 of galaxies with respect to the galaxy Main Sequence. In addition, \cite{Buat2009}, by using Spitzer MIPS data, also show that the dust attenuation
 expressed in terms of $log(L_{IR}/L_{UV})$ as a function of the $log(L_{IR}+L_{UV})$, which is proportional to the SFR, seems to be redshift
 independent (Figure 2 of \citealt{Buat2009}) in particular between redshift 0 and 1 as considered in this work. The same work also shows
 that $log(L_{IR}/L_{UV})$ as a function of the rest-frame K-band ($L_K$) luminosity, which is a proxy for the stellar mass, does
 not show any redshift dependence. This was recently confirmed also by \cite{Berta2013} with the most recent {\it{Herschel}} PEP
 and Hermes data. Thus, the substantial lack of evolution of IR and rest-frame UV properties of galaxies of a given mass and SFR,
 would suggest that the low redshift IR detected galaxies that populate the low star formation region of Figure \ref{fig2} can be
 used to calibrate the SFR$_{SED}$ of IR undetected galaxies in the same SFR region at higher redshift.

We point out that in the COSMOS field, as shown in the central panels of Figure \ref{fig2}, our calibration does not consistently move all
 galaxies to the 1 to 1 line (middle panel). High star-forming galaxies still show a slightly too low SFR$_{SED}$ with respect to the IR measure.
 This is probably due to the fact, that in the case of the \cite{Ilbert2010} SED fitting results, a plane in log-log space is not the best
 analytical form and, thus, it does not provide the best representation of the SFR$_{IR}-$SFR$_{SED}-$Mass relation. However, we still improve
 the agreement within $SFR_{SED}$ and $SFR_{IR}$ by more than a factor of two even in this field.
\subsection{The final galaxy group and group galaxy samples}
\label{sample}

The aim of our analysis is to study how the star formation activity in group-sized halos depends on the global
 properties of the systems.  In order to do that, we would need to sample the complete group galaxy population in
 stellar mass and SFR.  However, since the group members are spectroscopically selected, we need to consider how
 the spectroscopic selection function drives our galaxy selection and, thus, how it can affect our results. We point out that we can not define a galaxy sample which is, at the same time, complete in stellar mass and SFR.
For this purpose, we check how the spectroscopic completeness in the IRAC band translates into a completeness in mass and SFR.
For ECDFS and CFDN this is already done in \cite{Ziparo2013}. For the new datasets of AEGIS and COSMOS we follow the same approach of the mentioned work. This is done separately
 in two redshift bins ($0.15 < z < 0.5$ and $0.5 < z < 1.1$). The reference catalogs used to estimate this completeness
 are the photometric catalogs described in Section \ref{sed_fitting}. All those photometric catalogs are IRAC selected
 at 3.6 or 4.5 $\mu$m and should ensure photometric completeness down to at least $m_{AB}(3.6 {\mu}m) \sim 23$.
 From these catalogs we extract, for each field, the photometric redshift, the stellar mass and the SFR information
 derived from the SED fitting technique, after replacing the SFR$_{SED}$ with SFR$_{IR}$, where available, and after
 correcting SFR$_{SED}$ with the calibration presented in Section \ref{sed_fitting}.
Given the high accuracy of the photometric redshifts of \cite{Cardamone2010}, \cite{Wuyts2011}  and \cite{Ilbert2010},
 we assume the photometric redshifts, and the physical properties based on those, to be correct.
 We, then, estimate the completeness per stellar mass and SFR bins, respectively, as the ratio between the number of galaxies
 with spectroscopic redshift and the total number of galaxies in that bin. This procedure allows us to determine how the
 spectroscopic selection, based on the photometric information (e.g. colour, magnitude cuts, etc.), affects the choice of
 galaxies as spectroscopic targets according to their physical properties. Indeed,  Figure \ref{inco} illustrates while in any bin of stellar mass, the most star-forming galaxies are preferentially selected,
 the most massive galaxies are preferentially observed at any given SFR. 
 
  Thus, we follow the following approach to deal with spectroscopic incompleteness. We fix the stellar mass threshold
 to a value of 10$^{10}$ M$_{\odot}$, which guarantees a minimum spectroscopic completeness (40\%) for our analysis. We impose that this minimum completeness level (40\%) above the stellar mass threshold (10$^{10}$ M$_{\odot}$) must be reached in the region
 of the group. This completeness is estimated as follows. We consider
 a cylinder along the line of sight of the group with a radius of $2 \times r_{200}$ from the X-ray center and half
 width in redshift equal to 5$\times \sigma_{\Delta z/(1+z)}$, where $\sigma_{\Delta z/(1+z)}$ is the error of
 the photometric redshifts in each survey. This width is set to be much larger than the photometric
 redshift uncertainty and still small enough to sample the group region. The completeness is the ratio of the number of galaxies
 with spectroscopic redshift to the number of galaxies with spectroscopic or photometric redshift within this cylinder, with stellar mass above the given mass threshold. We perform the same analysis with different values of the
 cylinder half width (up to 10$\times \sigma_{\Delta z/(1+z)}$) and we obtain consistent measures of the completeness in mass. This minimum completeness level of 40\%  is fulfilled for almost all groups in the AEGIS, ECDFS, CDFN due to a very high and spatially homogeneous spectroscopic completeness. However, the 40 $\%$ threshold is hardly reached in many of the COSMOS group regions. The requirements is mainly fulfilled by the groups in the zCOSMOS region and by the GEEC2 groups. To deal with the reliability of our method, we analyse the possible biases induced by the spectroscopic selection
 function using mock catalogs. Our approach is explained in Section \ref{estimate}.


 We point out that the use of the full zCOSMOS and the GEEC2 spectroscopic sample increases the level of completeness in the COSMOS field
 by 20\% in the mean and in the group regions with respect to \cite{Ziparo2013}.



\begin{figure}
 \includegraphics[width=80mm]{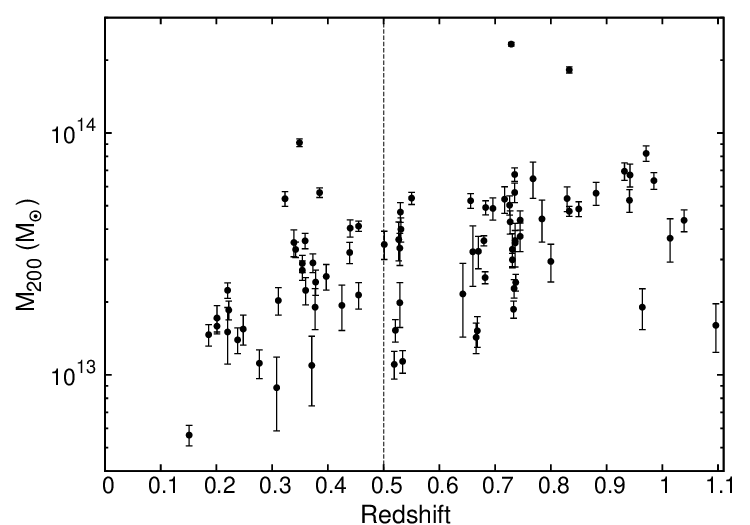} 
  \caption{M$_{200}$ vs. redshift for the final sample of galaxy groups in our analysis. The vertical dashed line separates low-z and high-z sample.} 
\label{fig4}
 \end{figure}

The final group sample is shown in Figure \ref{fig4}. The sample comprises 83 galaxy groups in the redshift range $0.15 < z < 1.1$.
 In order to study the evolution of the relation between the SF activity in groups and the system global properties, we divide
 the sample in two subsamples at $0.15 < z < 0.5$ (31 galaxy groups) and $0.5 < z < 1.1$ (52 galaxy groups). For 29 of 83 galaxy groups we have velocity dispersion from
 dynamical analysis and for the rest of them from X-ray properties. 50 of galaxy groups have Flag=1 and the remaining 33 have Flag=2. 

\subsection{The reference nearby group sample}
Our group sample does not cover the local Universe. Indeed, we apply a cut at $z=0.15$ in order to sample the same cosmic time epoch
 in the two redshift bins ($\sim 3$ Gyrs) considered in our analysis.  In order to follow the evolution of
 the group galaxy population down to $z\sim 0$, we complement our sample with a reference sample of nearby groups. Unfortunately,
 an X-ray selected sample of nearby groups in the same mass range as our sample with the same information as our groups, does not exist.
 Most of the X-ray selected samples available in the literature have a quite complicated selection function. In addition we
 need also a complete, spectroscopically confirmed, membership of any system and auxiliary information of the group galaxy
 stellar mass and star formation activity. Thus, we choose as a reference sample an optically-selected sample of nearby groups
 drawn from the SDSS and with a well studied and clean selection function. The group catalog and its general properties are
 discussed in \cite{Yang2007}. The catalog is drawn from the clean NYU-VAGC DR4 galaxy catalog (\citealt{Blanton2005}), which
 is a subsample of the SDSS DR4 galaxy spectroscopic catalog. The group selection is based on the halo-based group finder
 of \cite{Yang2005}, that is optimized for grouping galaxies that reside in the same dark matter halo. The performance
 of this group finder is extensively tested using mock galaxy redshift surveys constructed from the conditional luminosity
 function model (\citealt{Yang2003, vandenBosch2003, Yang2004}). The \cite{Yang2007} group
 catalog provides for each system the group membership and an estimate of the halo mass (M$_{200}$) (see \citealt{Yang2007} for a detailed discussion).
 In order to study the SF activity of nearby groups, we complement the group galaxy catalog of \cite{Yang2007} with the stellar
 masses and the SFR based on SDSS $H{\alpha}$ emission estimated by \cite{Brinchmann2004}. These quantities are corrected from aperture to total and to the
 same IMF used in our work. We also apply the same stellar mass cut ($M_* > 10^{10}$) and completeness level ($> 40\%$) in the nearby group sample for consistency.

\begin{figure}
\includegraphics[width=80mm]{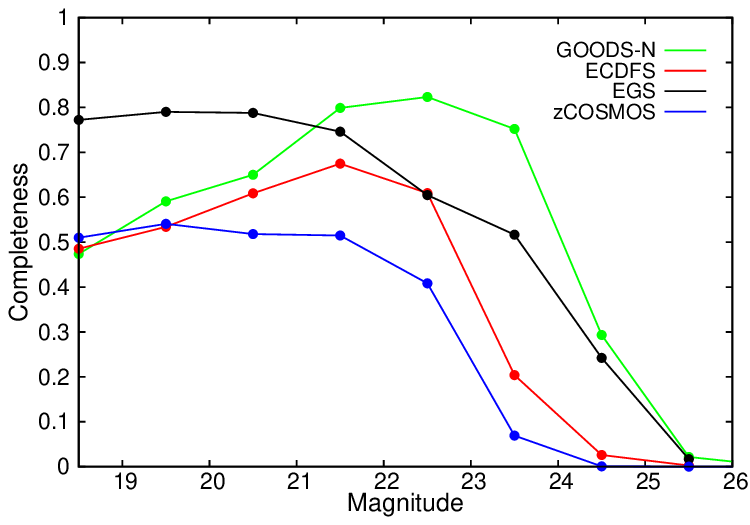}
\caption{Spectroscopic completeness for different fields in R-band magnitude. We use v-band magnitude for GOODS-N.}
\label{fig3}
\end{figure}

\subsection{The Millennnium mock catalogs \label{mock}}

\label{Millennium}

In order to estimate the errors involved in our analysis and check for possible biases
 due to the spectroscopic incompleteness, we follow the same approach used in \cite{Ziparo2013}
 based on the mock catalogs provided by the Millennium Simulation \citep{Springel2005}. The Millennium simulation follows the hierarchical growth of dark matter structures from redshift
 z = 127 to the present \citep{Springel2005}. Out of several mock catalogues created from the Millennium
 simulation, we choose to use those of \cite{Kitzbichler2007} based on the semi-analytical model of
 \cite{Delucia2006}. The simulation assumes the concordance $\Lambda$CDM cosmology and follows the
 trajectories of $21603^3\sim 1.0078\times10^{10}$ particles in a periodic box 500 Mpc h$^{-1}$ on a side.
  Kitzbichler \& White (2007) make mock observations of the artificial Universe by positioning a virtual
 observer at $z\sim0$ and finding the galaxies which lie on a backward light-cone. The backward light-cone
 is defined as the set of all light-like worldlines intersecting the position of the observer at redshift zero.
  We select as information from each catalogue the Johnson photometric band magnitudes available ($R_J$ , $I_J$ and $K_J$ ),
 the redshift, the stellar mass and the star formation rate of each galaxy with a cut at $I_J$ $<$ 26 to limit the data volume
 to the galaxy population of interest. In order to simulate the spectroscopic selection function of the surveys used in this work,
 we choose one of the available photometric bands ($R_J$) and extract randomly in each magnitude bin a percentage of galaxies consistent
 with the percentage of systems with spectroscopic redshift in the same magnitude bin observed in each of our surveys.
 We do this separately for each survey, since each field shows a different spectroscopic selection function as shown in
 Figure \ref{fig3}. We follow this procedure to extract randomly 25 catalogs for each survey from different light- cones.
 The ``incomplete'' mock catalogues, produced in this way tend to reproduce, to a level that we consider sufficient to our needs,
 the selection of massive and highly star-forming galaxies observed in the real galaxy samples, as already shown in \cite{Ziparo2013}.

We note that the galaxy mock catalogs of the Millennium simulation fail in reproducing the correct distribution of
 star-forming galaxies in the SFR-stellar mass plane, as already shown in \cite{Elbaz2007} at higher redshift ( $z \sim 1$),
 although they provide a rather good representation of the local Universe. This is caused by the difficulty of the semi-analytical models of predicting the observed evolution
of the galaxy stellar mass function and the cosmic star formation history of our Universe \citep{Kitzbichler2007, Guo2010}.
 We stress here that this does not produce a problem for our approach. Indeed, we aim to understand the bias induced by selection function like
the spectroscopic selection function of our dataset by using the Millennium galaxy mock catalogues.  In other words, we only need to extract mock
 catalogues randomly to reproduce the same bias in selecting, on average, the same percentage of most star-forming and most massive galaxies of the
 parent sample. By comparing the results obtained in the biased randomly extracted mock catalogues and the unbiased parent catalogue, we estimate
 the bias of our analysis. Since in both biased and unbiased mock catalogues the underestimation of the SFR or the stellar mass of high redshift galaxies
exists,  it does not affect the result of this comparative analysis. We also stress that the aim of this analysis is only to provide a way to interpret
 our results in terms of possible biases introduced by the spectroscopic selection function not to provide correction
factors for our observational results.

\section{Estimation of Total M$_*$ , Total SFR and Halo occupation Distribution of galaxy groups}
\label{estimate}

In this section we describe our method for estimating the total stellar mass ($\Sigma$M$_*$),
 the total star formation rate ($\Sigma$SFR) and the Halo Occupation Distribution of the galaxy
 groups in our sample. As explained in Section \ref{sample}, we impose a stellar mass
 cut at M$_* >$ 10$^{10}$M$_{\odot}$ since below this limit the spectroscopic completeness
 is rather low in all considered fields (see left panel of Figure \ref{inco}). The halo occupation distribution of each group, N(M$_*>$ 10$^{10}$ M$_{\odot}$),
 is defined by the number of galaxies with stellar mass above M$_*>$ 10$^{10}$M$_{\odot}$. The total stellar mass and star formation rate of
 each system are estimated as the sum of the group galaxy members stellar mass and SFR, respectively, with mass above the given limit. 
We correct for spectroscopic incompleteness by dividing $\Sigma$M$_*$, $\Sigma$SFR and N(M$_*>$ 10$^{10}$M$_{\odot}$) by the spectroscopic completeness estimated as explained in Section \ref{sample}.
In order to check if there are biases in our estimates due to the spectroscopic selection
function or to our method, and to calculate the uncertainties of each quantity, we use the
 galaxy mock catalogs described in Section \ref{Millennium}. For this purpose we extract from
 the original Kitzbichler \& White (2007) Mock catalog a sample of galaxy groups in the same
 mass and redshift range of the observed sample. We base our selection on the dark matter
 halo virial mass which, according to \cite{Delucia2006}, is consistent
 with the mass calculated within r$_{200}$, as in the observed group sample. The members of
 the groups are identified by the same Friends of Friends (FoF) identification number, defined
 according to the FoF algorithm described in \cite{Delucia2006}. We assume that the group galaxy members identified by the FoF algorithm,
 which takes into account also the real 3D spatial distribution of galaxies, are the correct (``true'') group members.
 The ``true'' velocity dispersion, $\Sigma$SFR, $\Sigma$M$_{*}$ and N are, thus, the one based on this membership. 

We apply, then, our method for calculating the membership, the velocity dispersion, total M$_*$ , total SFR and halo occupation 
distribution on the ``incomplete''  mock catalogs described in Section \ref{Millennium}, which include also the effect of the 
different spectroscopic selection functions. For each group we assume the coordinates of the central galaxy (the identification 
of central and satellite galaxies is provided in the mock catalog) as group center coordinates. These estimates are based on 
the 2D projected galaxy distribution and redshift information as in the real dataset. In this way we take into account both 
projection and incompleteness effects.  These quantities provide the ``observed'' velocity dispersion, $\Sigma$SFR, $\Sigma$M$_{*}$ and N.

\subsection{Reliability of group membership and velocity dispersion estimate}

\begin{figure}
\centering
\includegraphics[width=0.39\textwidth,angle=0]{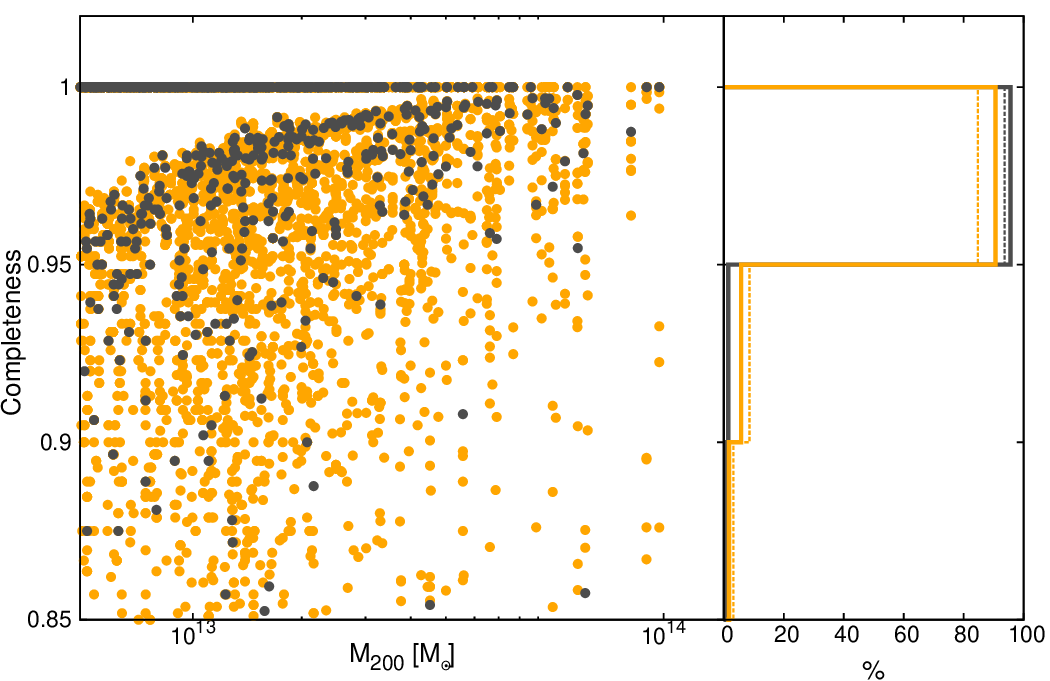}
\includegraphics[width=0.39\textwidth,angle=0]{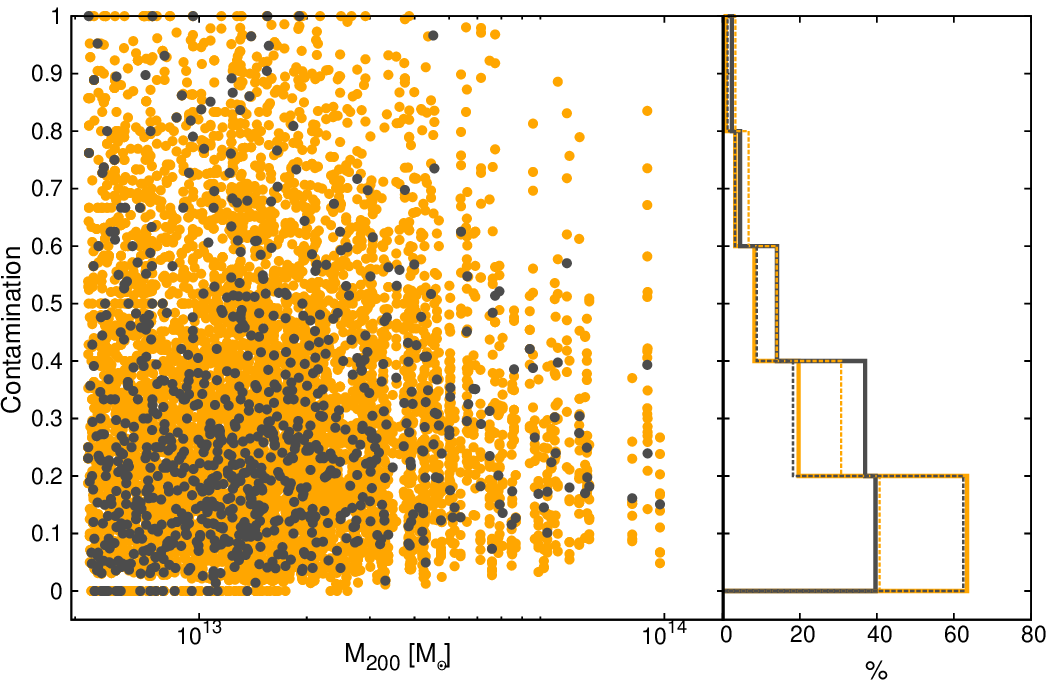}
\caption{Completeness and contamination level of the member galaxies using the gapper estimator method with initial condition from M$_{200}$
 (grey points) and M$_{200}$ with error (orange points) in the mock catalog. The right panels show corresponding histograms. The solid lines
 in the histograms show galaxies with M$_*$ $>$ 10$^{10}$M$_{\odot}$ and the dashed histograms are related to the whole sample.}
\label{sigma2}
\end{figure}

\begin{figure}
\centering
\includegraphics[width=0.41\textwidth,angle=0]{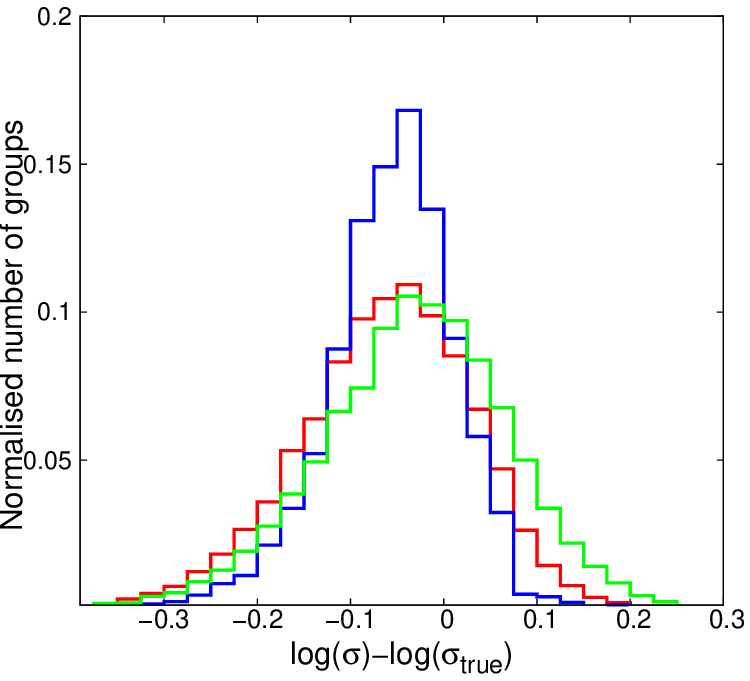}
\caption{Distribution of the residuals of the logarithm of the ``true'' and ``observed'' velocity dispersion.
 The blue histogram shows the distribution of the residuals obtained from the original mock catalogs. The red
 histogram shows the distribution obtained if we take into account the error on M$_{200}$ derived from $L_X$
 as done in the real dataset. The green histogram show the same diagram but with the ``observed'' velocity
 dispersion estimated on the basis of the ``incomplete'' mock catalog. }
\label{sigma}
\end{figure}

In order to check if our method is able to recover efficiently the membership of each group, we compare the completeness and the
 contamination of the membership obtained in our analysis with the original group membership identified by the FoF algorithm of the mock catalog.
 The completeness is estimated by computing the fraction of ``true'' members identified by our method. The contamination is estimated by calculating
 the fraction of interlopers (galaxy identified as group members by our method but not in the original mock catalog). Figure \ref{sigma2} shows the completeness level (top panel) and the contamination level (bottom panel) of our group membership.
 The dashed histograms in both panels show the completeness and contamination levels obtained if we considered all members without any stellar
 mass cut. The completeness level is quite high ($> 90\%$) but on average 35\% of the members are interlopers. If we apply a mass cut
 of $10^{10}M_{\odot}$, the completeness level reaches almost in all cases 100\% with a much lower contamination fraction (solid histograms).
 It is clear that our method is much more robust in identifying rather massive galaxy members, which are likely more clustered in the
 phase space, than low mass galaxies. The red and blue histograms (Figure \ref{sigma}) indicate the cases in which the velocity dispersion first guess
 is estimated from the mock catalog M$_{200}$ without and with error, respectively (see below). After performing the same recovery test on
the ``incomplete'' mock catalog, we check that the completeness level is driven by the mean simulated completeness of the sample,
while the contamination level remains at the same values.

We estimate the ``observed'' velocity dispersion on the basis of this membership to take into account the effect of spectroscopic incompleteness.
 We measure the ``observed'' $\sigma$ as in the real dataset. In other words, we base the velocity dispersion estimate on M$_{200}$ and the
 relation between $\sigma$ and r$_{200}$ as in \cite{Carlberg1997} for groups with less
 than 10 members and on the dynamical analysis for groups with more than 10 galaxies. We consider also that our first guess for the velocity dispersion is affected by the
 uncertainty in the M$_{200}$ in the observed dataset, which is retrieved via L$_X-$M$_{200}$ correlation. To take this into
 account we add a random error to the M$_{200}$ of the group provided by the mock catalog. The scatter
 of the L$_X-$M$_{200}$ relation is quoted about 20\% in the group mass region based on the estimation via stacking analysis (\citealt{Leauthaud2010, Allevato2012}).
 However, to be conservative, we use the L$_X-$T$_X$ relation and scatter reported in Sun (2011) to estimate a scatter in the L$_X-$M$_{200}$ relation.
 We use a value of 0.3 dex in our exercise. The green histogram of Figure \ref{sigma} shows the residual distribution between the ``true'' and
``observed'' velocity dispersion. The two values are in rather good agreement with a scatter of 0.1 dex. The main source of scatter is given
by the spectroscopic incompleteness. Indeed, if we perform the same test by using the original ``complete'' mock catalog, the scatter
decreases to 0.06 dex (blue histogram) and it is due to projection effects. The uncertainty in the first guess of the velocity dispersion does not affect significantly the
final estimate. Indeed, without including this source of error the scatter decreases only to 0.09 dex (red histogram).

As shown in Figure \ref{sigma}, the peak of the
residual distribution is not zero but it shows that we tend to underestimate the true velocity dispersion
 by $\sim$20\%. This shows that the \cite{Carlberg1997} relation (used for estimating the first guess, in general, and the
 velocity dispersion for systems with less than 10 members, in particular) is not itself a source of scatter but it could cause
 a bias in the estimation of velocity dispersion.

\begin{figure}
\centering
\includegraphics[width=0.39\textwidth,angle=0]{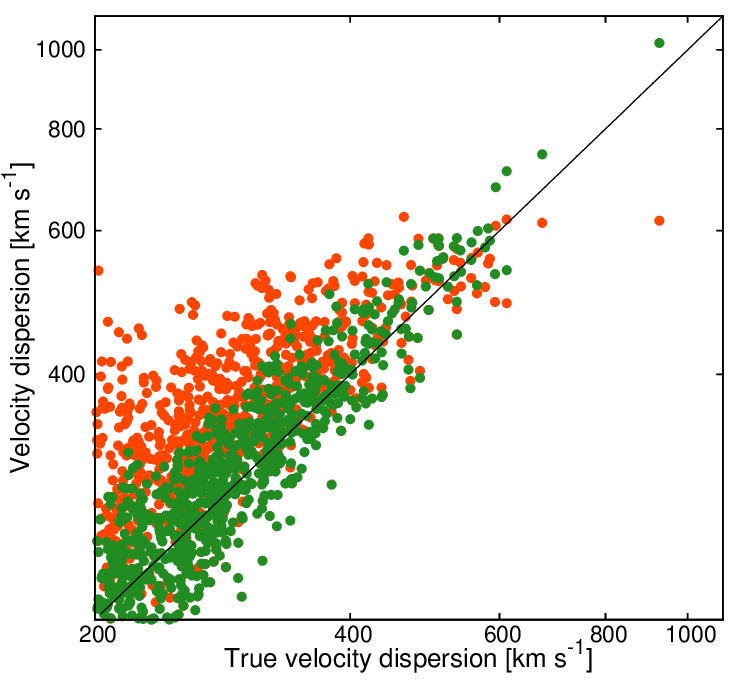}
\caption{ Velocity dispersion from gapper estimator vs. true velocity dispersion for mock groups. The orange points show the choice
 of constant initial velocity dispersion and the green one is based on the initial velocity dispersion computed from M$_{200}$}
\label{sigma1}
\end{figure}

We also point out that using the estimate of M$_{200}$ for deriving the
 velocity dispersion first guess is a fundamental ingredient of our analysis. Indeed, if we use a constant value for the first
 guess, as usually done in the literature, we find that the scatter in the relation between ``true'' and ``observed'' velocity dispersion increased significantly
 as shown in Figure \ref{sigma1} (orange points) and there is no good correlation between the two quantities.

\subsection{Reliability of Total M$_*$ , Total SFR and Halo Occupation Distribution}
\label{error}
\begin{figure*}
\includegraphics[width=0.3\textwidth,angle=0]{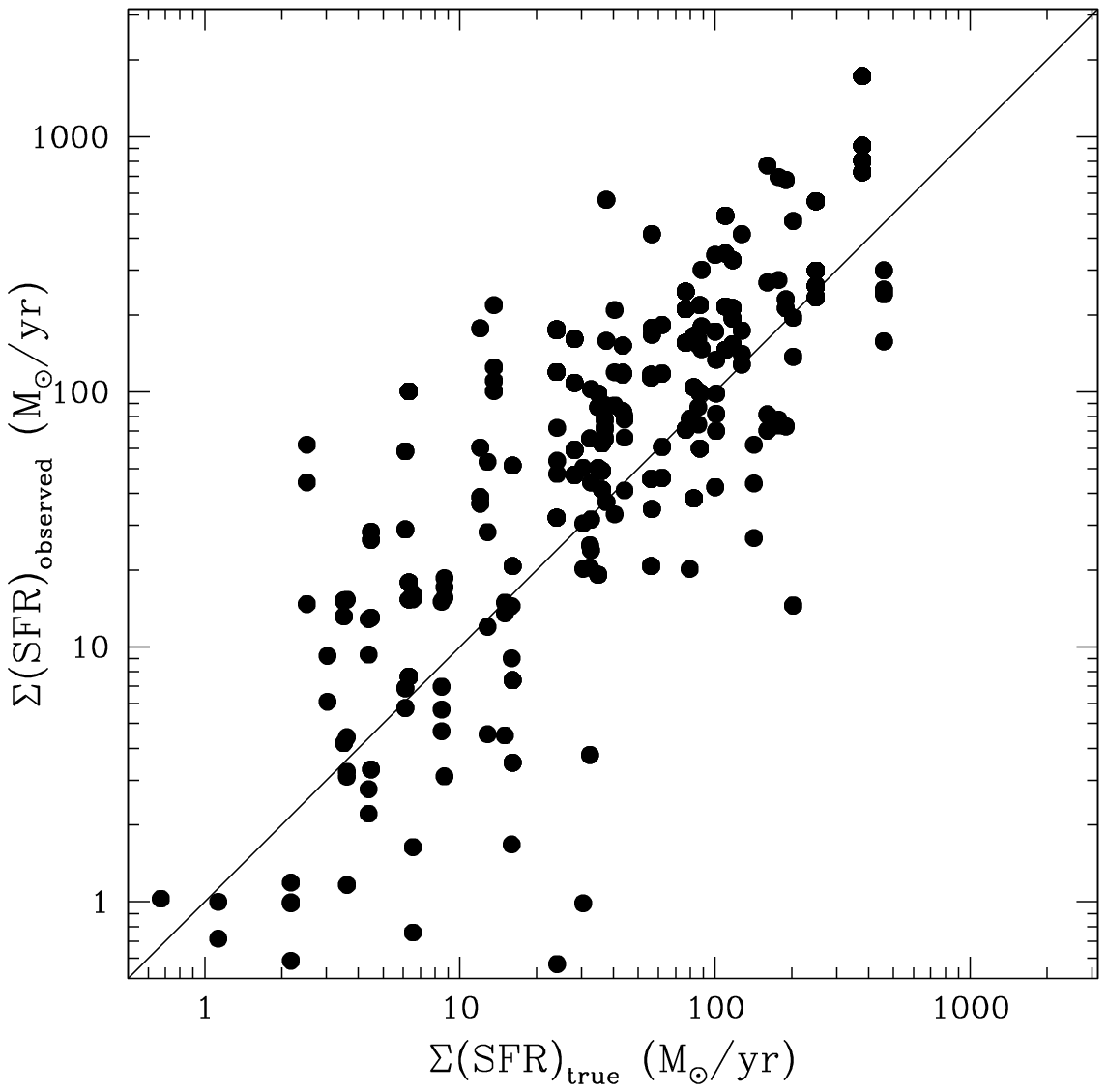}
\includegraphics[width=0.3\textwidth,angle=0]{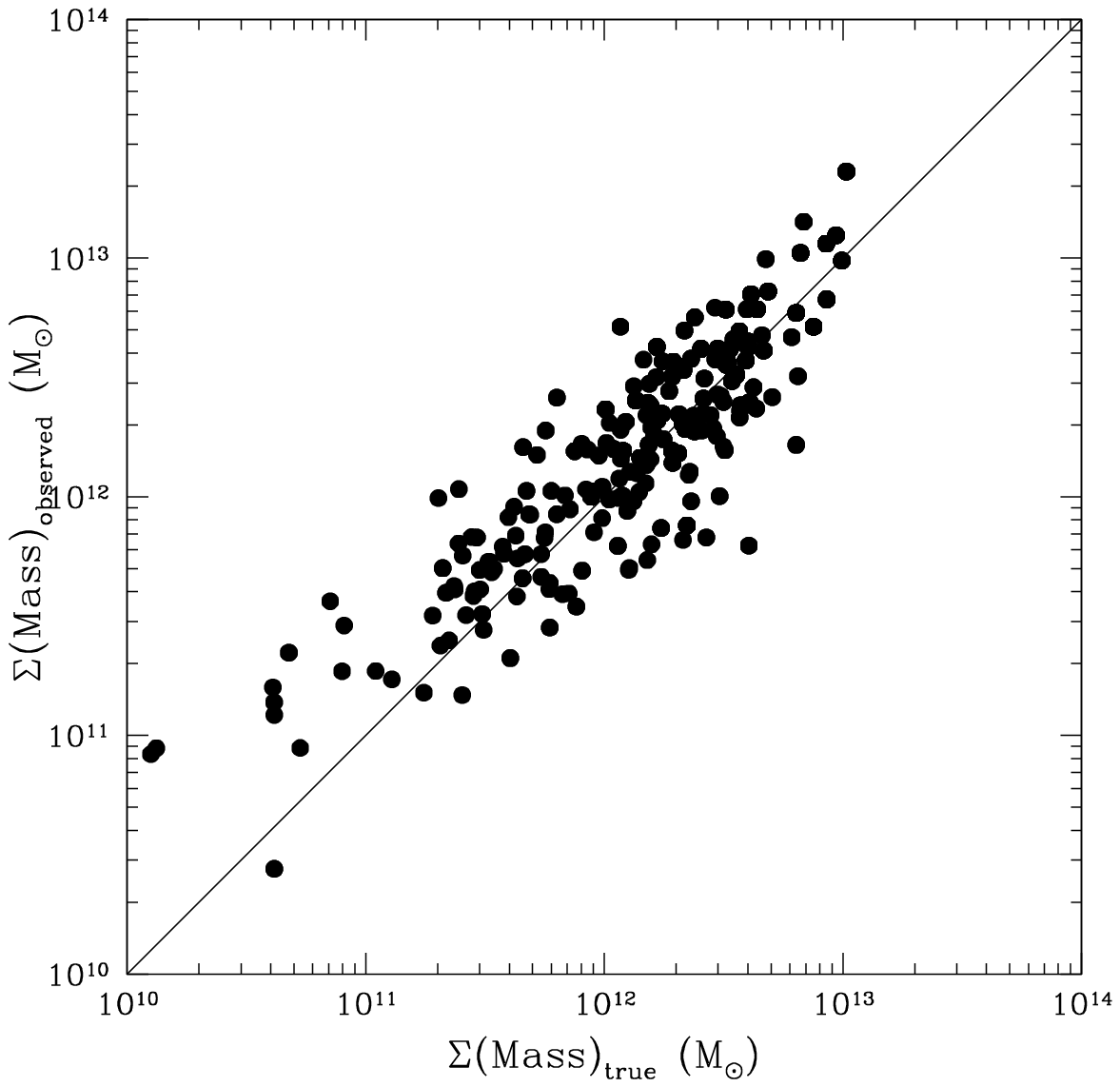}
\includegraphics[width=0.3\textwidth,angle=0]{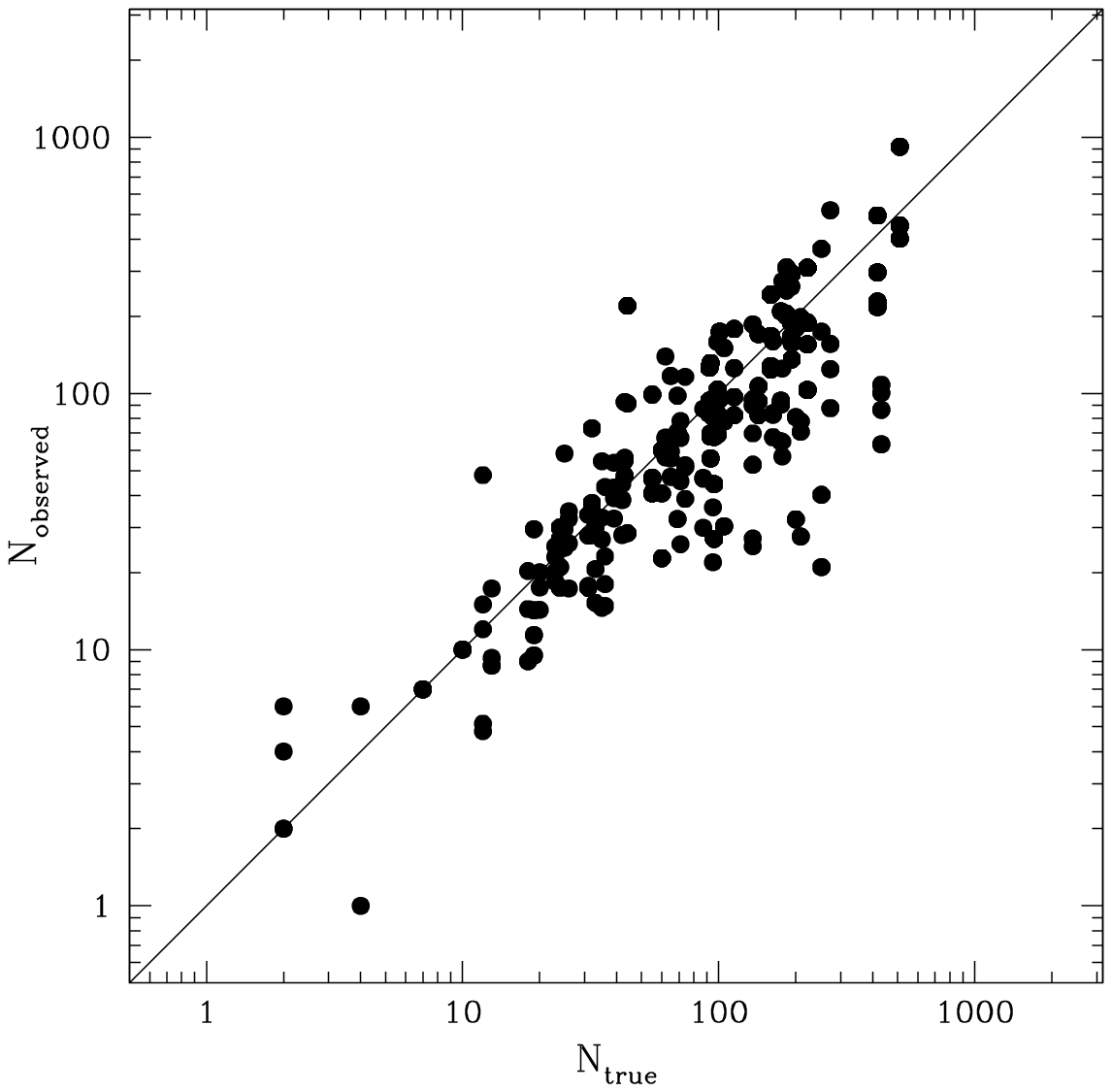}
\caption{ From left to right, ``True'' values of total SFR, total stellar masses and halo occupation number of the groups vs. our estimates in the ``incomplete catalogs'' with the same level of spectroscopic incompleteness
 of the surveys used in this work.}
\label{incomplete}
\end{figure*}

As for the ``observed'' velocity dispersion, we also estimate the ``observed'' total stellar mass,
 total star formation rate and halo occupation distribution by applying our method to the ``incomplete'' mock catalogs
 to include the effect of projection and spectroscopic incompleteness. Each estimate is obtained after applying our
 stellar mass cut at M$_*>$ 10$^{10}$M$_{\odot}$.  We also apply the correction for incompleteness as described in
 Section \ref{sample}. Figure \ref{incomplete} shows the comparison of the ``true'' and ``observed" quantities.
 We find a rather good agreement between the two values in all cases. However, we notice a large scatter (0.3 dex) between
 the ``true" and ``observed" total SFR and a smaller scatter for ``true" versus ``observed" total M$_{*}$ (0.17 dex) and N (0.15 dex).
 This different behavior of the scatter is due to two aspects. On average, the galaxies contaminating the group membership are field galaxies,
 likely less massive, due to mass segregation ( less massive galaxies prefer low density regions while more massive galaxies mostly located in high density environments, \citealt{Scodeggio2009}), and more star-forming than group galaxies. This is true in particular for the Millennium Simulation mock
 catalogs that are affected by an overabundance of red and dead galaxies in groups due to the satellite overquenching problem described
 in \cite{Weinmann2009}. The result of this overquenching is that the level of the star formation in group galaxies
 is suppressed with respect to less crowded environments. Thus, in the case of groups with a low number of galaxies, the presence of even one
 contaminant with a high star formation rate can highly alter the total level of star formation activity. On the other hand, group galaxies tend
 to be rather massive and the addition of one or few field galaxies of average mass does not much affect the total M$_{*}$ of the system.
 Thus, the uncertainty turns out to be much larger in the total SFR than in the total M$_{*}$ or the N. Since in the local Universe we do not
 observe such a high abundance of red and dead satellite in groups as in the mocks (\citealt{Weinmann2009}), it is likely that the uncertainty from the total SFR estimated
 in the Millennium Simulation mock catalogs is overestimating the actual uncertainty. 

The low level contamination (see previous section) also explains why in some cases we observed a slightly larger number of galaxies in groups with respect to the ``true" value.

\section{Results}

In this section we analyse several relations. First we study the correlation between the total SFR in groups
 versus the group global properties such as L$_X$, $\sigma_v$ and M$_{200}$. Since L$_X$ and $\sigma_v$ are the only independent measurments and they also exhibit a relation with a tight scatter (Figure \ref{fig1}),
 we discuss in particular only the $\Sigma$SFR$-$M$_{200}$ relation to
 relate the evolution of the star formation activity of the group population to the total DM halo mass. However, all the relations
 derived are listed in Table \ref{fitMN}. As previously mentioned, we divide our sample in two \textquotedblleft low\textquotedblright
 and \textquotedblleft high\textquotedblright redshift bins ($0.15<z<0.5$ and $0.5<z<1.1$). The two redshift bins are defined in order
 to have enough statistics and to sample a comparable fraction of the age of the Universe ($\sim3$ Gyr) at different epochs. However, we must take into account
 that the two bins are rather wide and a significant evolution in terms of stellar population can occur in galaxies in such a large amount of time. 
 In the low redshift bin, a Spearman test confirms that in none of the considered cases there is a significant correlation, while there is a rather poor correlation
 between the total mass M$_{200}$ and the group redshift as already visible in Figure \ref{fig4}. Thus, in the low redshift bin, the different evolutionary state of the galaxy population
 of groups can be an additional source of scatter in the analysed relations but it does not affect the slope of the relation.
 
However, at high redshift we observe a quite significant correlation between each quantity and the redshift. These correlations are induced by the
 strong correlation between M$_{200}$ and the group redshift as visible in Figure \ref{fig4} at $z > 0.5$. This correlation is due to the X-ray selection
 that tends to select higher mass systems at high redshift. In order to take this selection effect into account we select a subsample of the high
 redshift groups in the redshift range $0.5 < z < 0.8$. This subsample comprises 38 systems and it does not show any correlation between M$_{200}$, $\Sigma$SFR or $\Sigma$M$_{*}$ or N
 with the group redshift. We use this subsample to check whether the observed correlations between the aforementioned quantities  and their slopes are driven by a redshift dependence.

We perform the analysis of each correlation by estimating the quantities within $r_{200}$ and 2 $\times$ $r_{200}$. The results obtained
 within $r_{200}$ are consistent with the corresponding results in 2 $\times$ $r_{200}$. We present in this section the results obtained within 2 $\times$ $r_{200}$
 since this is the case with the best statistics.

\begin{table*}
\begin{tabular}{cccccc}

relation&z&Intercept  &Slope & Spearman $\rho$& Spearman P \\

\hline
\hline
log(M$_{200}$)-log($\Sigma$ SFR)&0.15-0.5& -7.68$^{\pm2.32}$&0.68$^{\pm0.17}$&0.3&0.02\\[4pt]
log(M$_{200}$)-log($\Sigma$ SFR)&0.5-1.1&-11.32$^{\pm1.52}$&1.00$^{\pm0.11}$&0.44&4e-6\\[4pt]

log(L$_X$)-log($\Sigma$ SFR)&0.15-0.5& -14.35$^{\pm5.9}$&0.37$^{\pm0.14}$&0.29&0.02\\[4pt]
log(L$_X$)-log($\Sigma$ SFR)&0.5-1.1&-23.22$^{\pm3.9}$&0.59$^{\pm0.09}$&0.47&3e-7\\[4pt]

log($\sigma$)-log($\Sigma$ SFR)&0.15-0.5& -1.32$^{\pm1.69}$&1.12$^{\pm0.5}$&0.26&0.02\\[4pt]
log($\sigma$)-log($\Sigma$ SFR)&0.5-1.1&-2.60$^{\pm1.00}$&1.93$^{\pm0.4}$&0.4&6e-5\\[4pt]

log(M$_{200}$)-log($\Sigma$ M$_*$)&0.15-0.5& -1.82$^{\pm3.23}$&1.02$^{\pm0.24}$&0.5&2e-4\\[4pt]
log(M$_{200}$)-log($\Sigma$ M$_*$)&0.5-1.1&-1.52$^{\pm3.67}$&0.99$^{\pm0.25}$&0.4&1e-5\\[4pt]

log(L$_X$)-log($\Sigma$ M$_*$)&0.15-0.5& -14.36$^{\pm4.53}$&0.62$^{\pm0.09}$&0.52&8e-5\\[4pt]
log(L$_X$)-log($\Sigma$ M$_*$)&0.5-1.1&-11.63$^{\pm4.5}$&0.55$^{\pm0.11}$&0.38&7e-5\\[4pt]

log($\sigma$)-log($\Sigma$ M$_*$)&0.15-0.5& 7.09$^{\pm0.93}$&1.95$^{\pm0.38}$&0.47&1e-4\\[4pt]
log($\sigma$)-log($\Sigma$ M$_*$)&0.5-1.1&6.88$^{\pm1.06}$&2.02$^{\pm0.42}$&0.37&8e-5\\[4pt]

log(M$_{200}$)-log(N)&0.15-0.5& -8.04$^{\pm1.98}$&0.67$^{\pm0.14}$&0.5&1e-4\\[4pt]
log(M$_{200}$)-log(N)&0.5-1.1&-10.87$^{\pm1.52}$&0.90$^{\pm0.11}$&0.57&1e-8\\[4pt]

log(L$_X$)-log(N)&0.15-0.5& -17.13$^{\pm3.65}$&0.43$^{\pm0.08}$&0.5&5e-4\\[4pt]
log(L$_X$)-log(N)&0.5-1.1&-21.39$^{\pm2.9}$&0.52$^{\pm0.06}$&0.51&0\\[4pt]

log($\sigma$)-log(N)&0.15-0.5& -2.27$^{\pm0.73}$&1.34$^{\pm0.31}$&0.44&1e-4\\[4pt]
log($\sigma$)-log(N)&0.5-1.1&-3.39$^{\pm0.75}$&1.81$^{\pm0.3}$&0.43&1e-6\\[4pt]

log(M$_{200}$)-SF fraction&0.15-0.5& 1.97$^{\pm4.08}$&-0.11$^{\pm0.3}$&-0.25&0.35\\[4pt]
log(M$_{200}$)-SF fraction&0.5-1.1&6.94$^{\pm1.9}$&-0.45$^{\pm0.13}$&-0.49&0.002\\[4pt]

log(L$_X$)-SF fraction&0.15-0.5& 2.40$^{\pm8.02}$&-0.045$^{\pm0.18}$&-0.251&0.34\\[4pt]
log(L$_X$)-SF fraction&0.5-1.1&13.3$^{\pm3.54}$&-0.29$^{\pm0.08}$&-0.5&0.001\\[4pt]

log($\sigma$)-SF fraction&0.15-0.5& 1.31$^{\pm1.0}$&-0.33$^{\pm0.41}$&-0.25&0.34\\[4pt]
log($\sigma$)-SF fraction&0.5-1.1&2.94$^{\pm0.66}$&-0.87$^{\pm0.26}$&-0.41&0.0097\\[4pt]



\hline\\
\end{tabular}
\caption{The table present all the best fit results of the ordinary least squares regression method performed on the low and high galaxy group sample. The first column indicates the considered $x-y$ relation.
 The second column indicates the redshift bin. The third and fourth columns indicate the intercept and the slope, respectively, of the best fit so
 that $y=slope*x+intercept$. The fifth column indicates the Spearman correlation coefficient and the sixth column indicate the value of the probability of the null
 hypothesis of no correlation among the considered quantities.}
\label{fitMN}
\end{table*}

\subsection{$\Sigma$ SFR, $\Sigma$ M$_*$ vs M$_{200}$ and N}

\begin{figure}
\centering
\includegraphics[width=70mm]{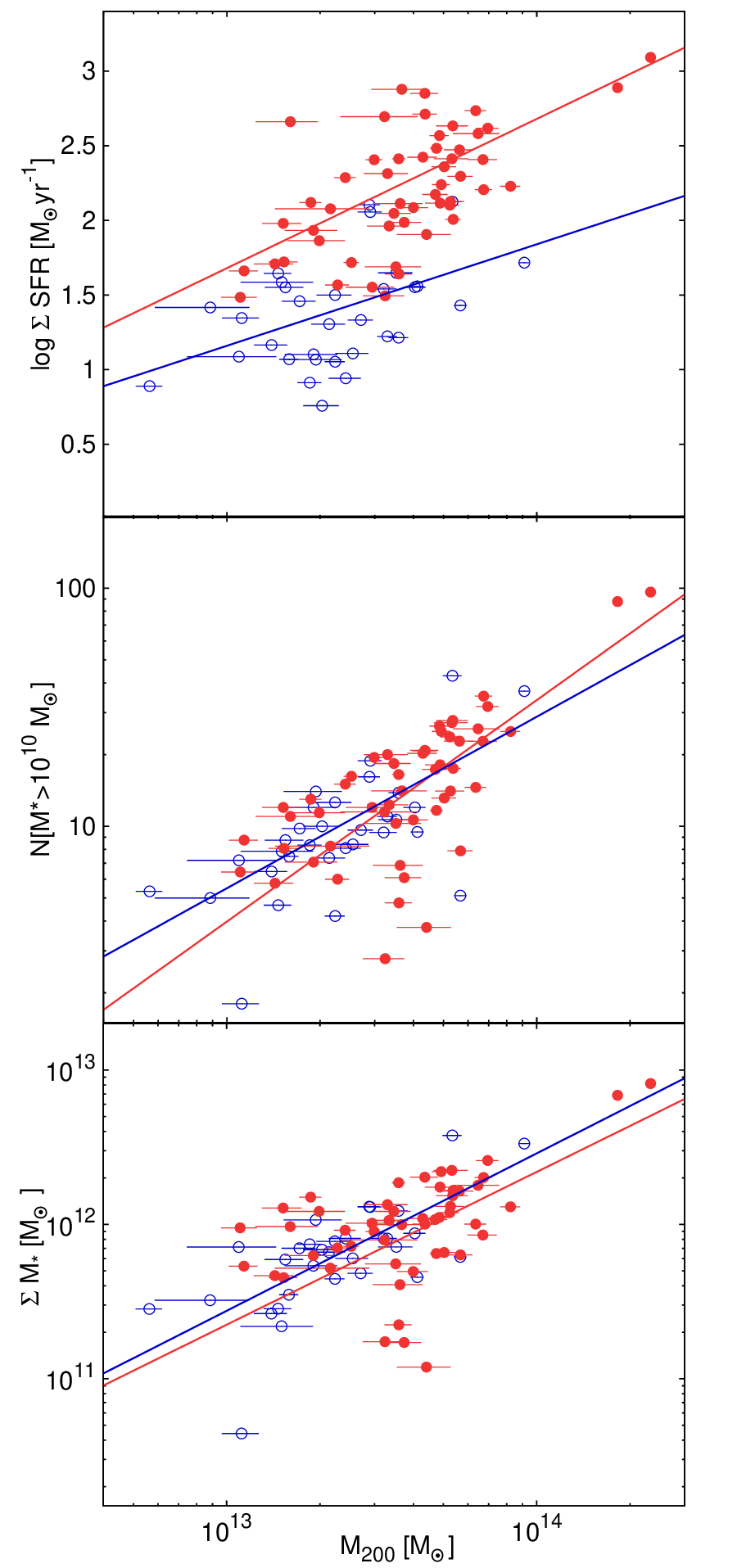}
 \caption{$\Sigma$ SFR- (upper panel), N- (middle panel) and $\Sigma$ M$_*$- (bottom panel)  M$_{200}$ relations for member
 galaxies with M$_*$ $>$ 10$^{10}$ M$_{\odot}$ in the low-z sample (0.15$<$z$<$0.5, in blue) and the high-z groups (0.5$<$z$<$1.1, in red).
The blue and red lines show the best-fitting using the ordinary least squares regression method presented by Akritas \& Bershady (1996). 
 The total star formation activity in high-z groups is higher with respect to the low-z  sample at any mass by $0.8\pm0.12$ dex.
 The N- and $\Sigma$ M$_*$- M$_{200}$ are consistent with a linear relation in both redshift bins with no evolution since z$\sim$1.1.}
\label{SFR}
 \end{figure}

The upper panel of Figure \ref{SFR} shows the $\Sigma$SFR-M$_{200}$ relation in the low (blue points) and high (red points)
 redshift bins. A Spearman correlation test shows a much more significant positive correlation in the high-z sample and a very mild correlation
 in the low-z one (see Table \ref{fitMN}).

\begin{figure*}
 \includegraphics[width=0.329\textwidth,angle=0]{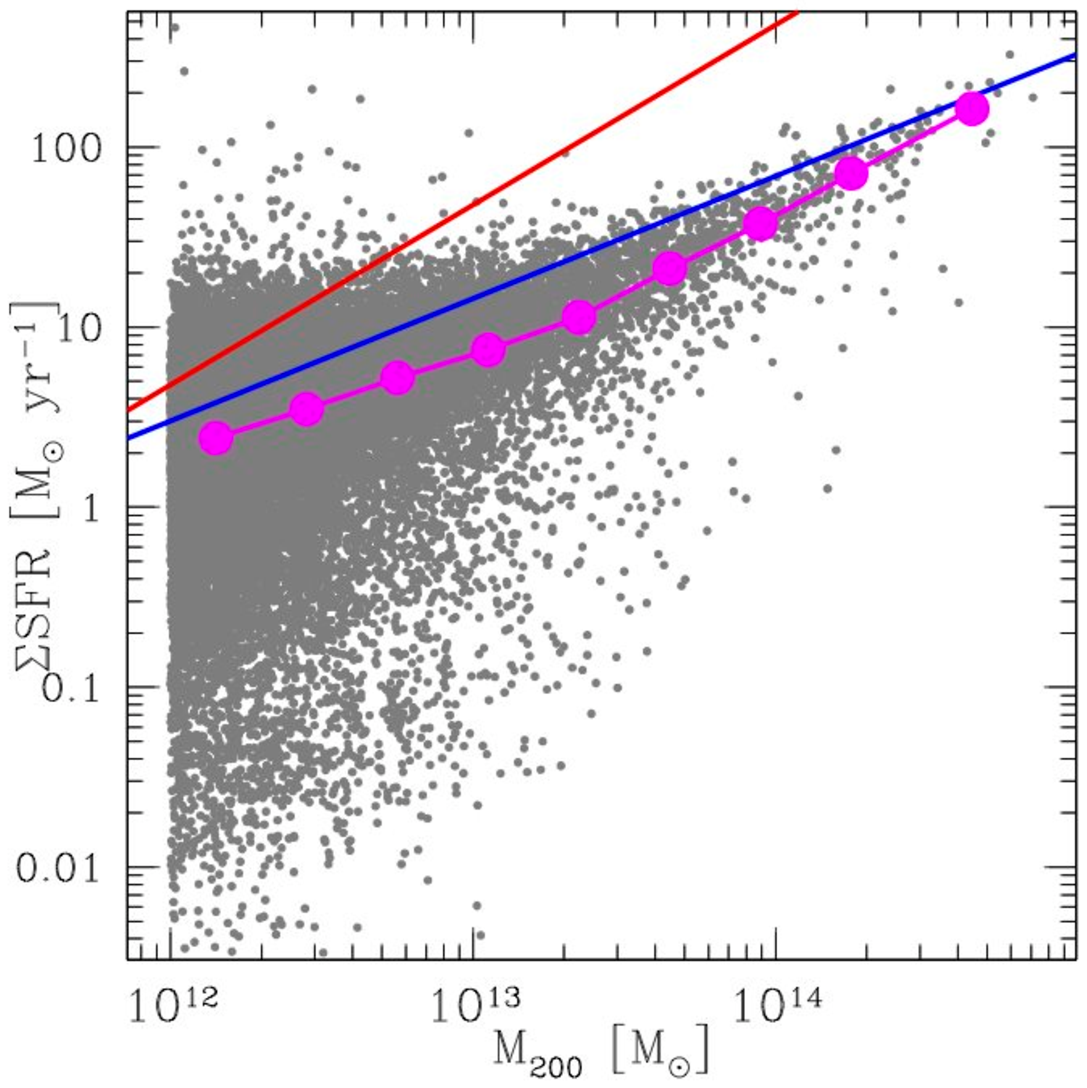}
\includegraphics[width=0.329\textwidth,angle=0]{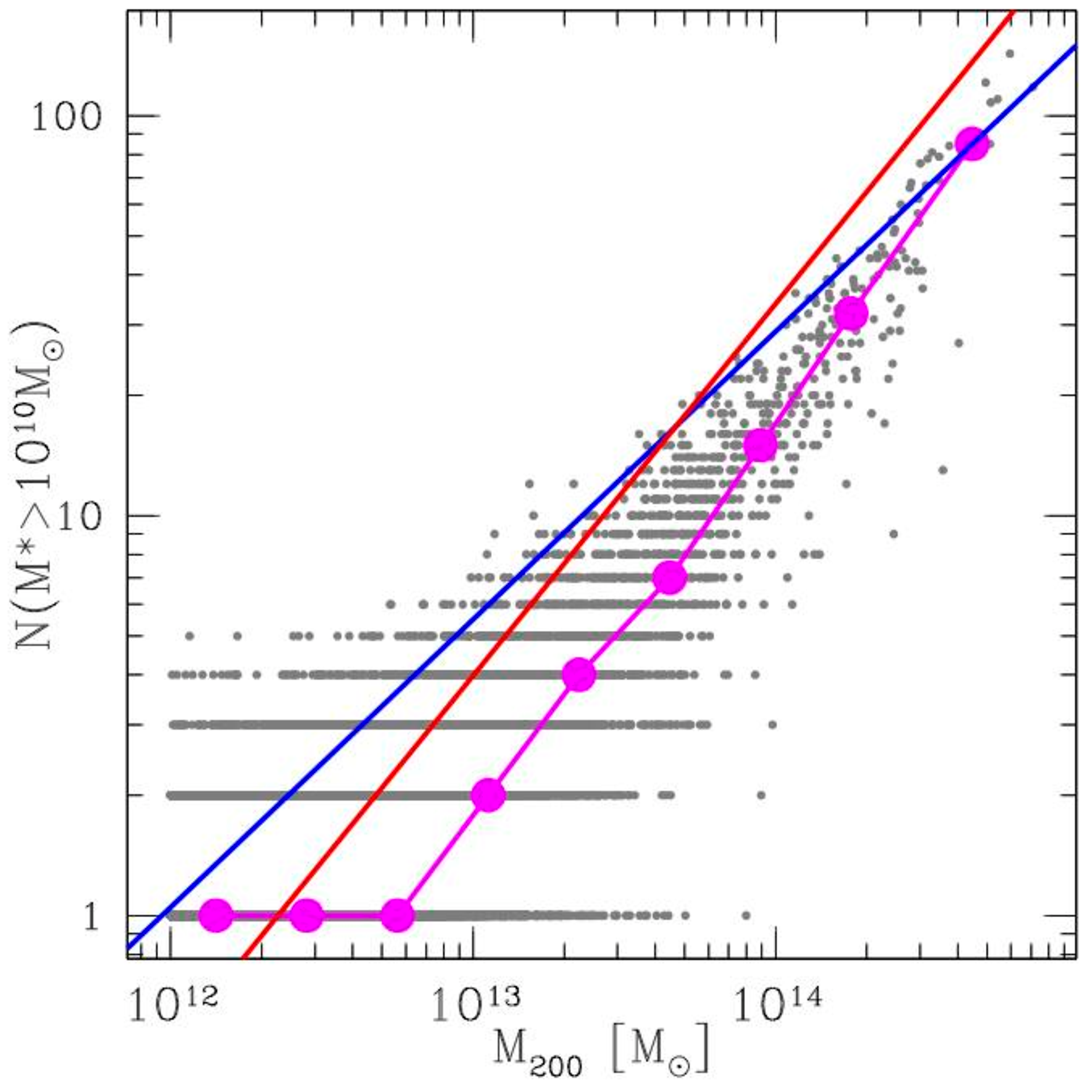}
\includegraphics[width=0.329\textwidth,angle=0]{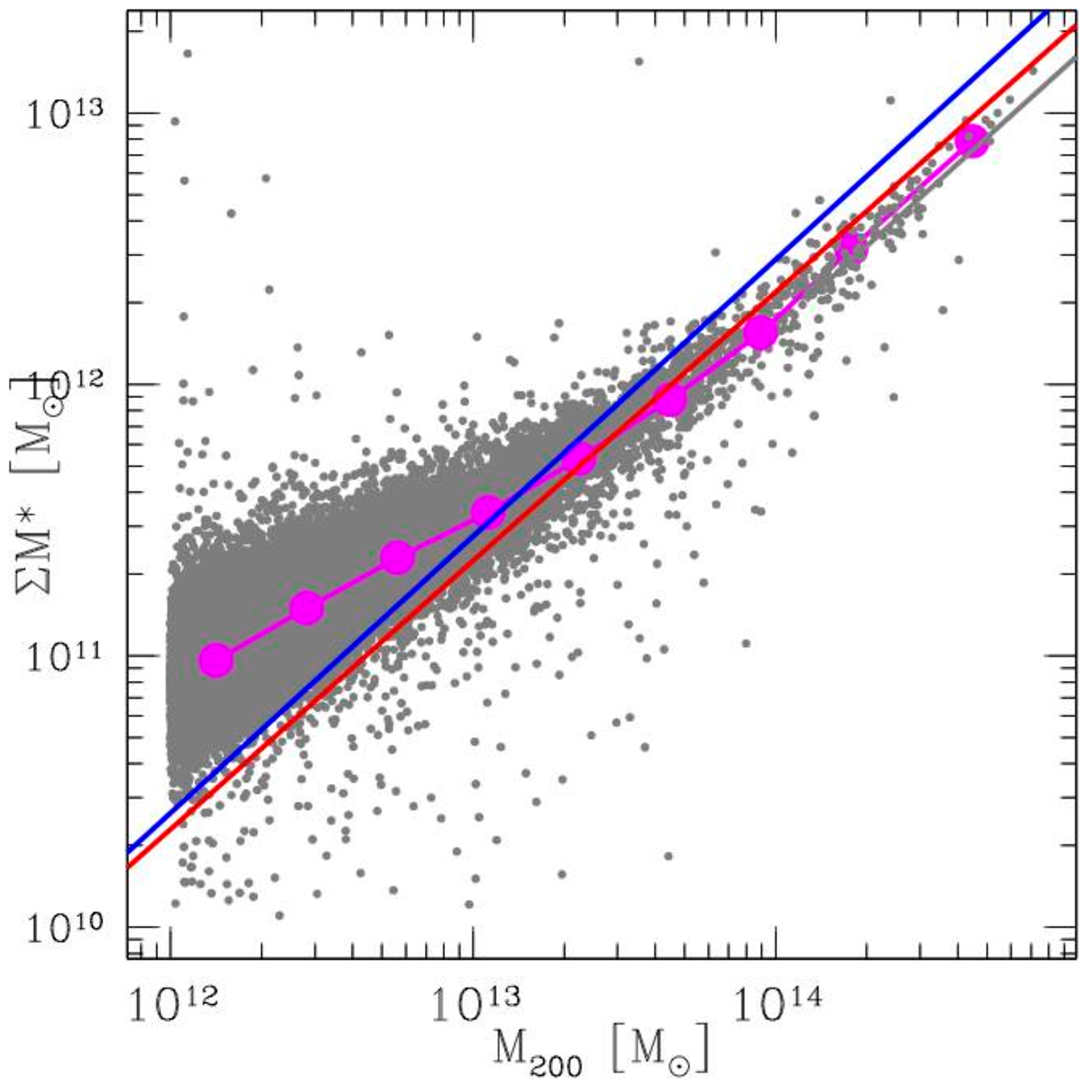}
 \caption{$\Sigma$ SFR- (left panel), N- (middle panel) and $\Sigma$ M$_*$- (right panel)  M$_{200}$ relations for a subsample of \citep{Yang2007} optically
 selected catalog at $z<0.085$ (grey points). The magenta points connected by the solid line shows the median per bin of M$_{200}$ in
 the \citep{Yang2007} subsample. The blue solid lines show the best fit relation of our low-z sample and the red solid lines show the best
 fit relation of our high-z group sample. The $\Sigma$ SFR and total mass of the nearby groups are strongly correlated. We do not see, however,
 a simple linear correlation in the log-log space but a double slope, flatter ($\Sigma{SFR} \propto M_{200}^{0.56\pm0.01}$) at M$_{200} < 10^{13}$M$_{\odot}$ and steeper
  ($\Sigma{SFR} \propto M_{200}^{0.89\pm0.03}$) at $M_{200} > 10^{13}M_{\odot}$.}
\label{yang}
 \end{figure*}

We first investigate the possibility that the lack of a significant correlation in the low redshift bin could be due to the low number statistics.
 Indeed, the low redshift bin contains 31 galaxy groups. This relatively low number together with the scatter due to the differences
 in the age of the stellar population of the group galaxies in such a wide redshift bin ($\sim$ 3Gyrs), could prevent us from observing a correlation.
 To check this possibility we use as a reference sample of nearby groups the optically-selected group sample of
 \cite{Yang2007} drawn from the SDSS. We select in particular a subsample of groups at $z < 0.085$. This is done because the
 SDSS spectroscopic sample is complete at masses $>$10$^{10}$M$_{\odot}$ below this redshift limit (see \citealt{Peng2010}).
 As shown in the left panel of Figure \ref{yang}, the total SFR and total mass of the nearby groups are strongly correlated. We do not see, however,
 a simple linear correlation in the log-log space but a double slope, flatter ($\Sigma{SFR} \propto M_{200}^{0.56\pm0.01}$) at M$_{200} < 10^{13}M_{\odot}$ and steeper
  ($\Sigma{SFR} \propto M_{200}^{0.89\pm0.03}$) at M$_{200} > 10^{13}$M$_{\odot}$. As explained by \cite{Yang2008} the break at the low-mass
 end can be explained by the Halo Occupation Statistics. Indeed, we observe the same sharp break in the N of the \cite{Yang2007} subsample
 at N(M$>$ 10$^{10}$M$_{\odot}$)$\sim$1 (central panel of Figure \ref{yang}). This break indicates that, on average, below M$_{200} \sim 10^{13}$M$_{\odot}$ only
 the central galaxy has a mass above M$>$ 10$^{10}$M$_{\odot}$ and satellites have lower masses. The existence of a significant correlation between $\Sigma$SFR and M$_{200}$ in the nearby groups and in the more populated high redshift group sample would suggest that we should likely observe a correlation also in the low redshift bin.
 Thus, to check if the low number statistics and the scatter are hiding such a correlation, we perform the following test. We extract randomly 5000 times the same number of objects as in the intermediate redshift sample from the \cite{Yang2007} subsample in the same mass range. We perform for each extraction the Spearman test between $\Sigma$SFR and M$_{200}$. In 65\% of the cases we observe a correlation between the two quantities of the same significance as in our
 low redshift sample. Thus, we conclude that the mild correlation observed in our low redshift group sample is due to low number statistics in
 addition to the scatter due to the width of the redshift bin.

To further check if the positive correlation between M$_{200}$ and the redshift of the groups in the high redshift bin
 can induce the positive correlation observed between $\Sigma$SFR and M$_{200}$, we consider the subsample of the high-z groups, described above,  at $0.5 < z < 0.8$. We perform the Spearman test and the ordinary least squares regression method (\citealt{Akritas1996}) in the log-log space of  $\Sigma$SFR and M$_{200}$ for such subsample and we find a correlation significance
 and slope to be perfectly consistent (within 1$\sigma$) with the results obtained with the whole high redshift sample. The effect of the addition of the remaining $z > 0.8$ groups is only to increase the scatter of the relation by 17\%.
Thus, we conclude that the positive correlation is not induced by a redshift bias in our group sample and that the positive correlation of the $\Sigma$SFR$-$M$_{200}$ relation is real.

\begin{figure}
\includegraphics[width=80mm]{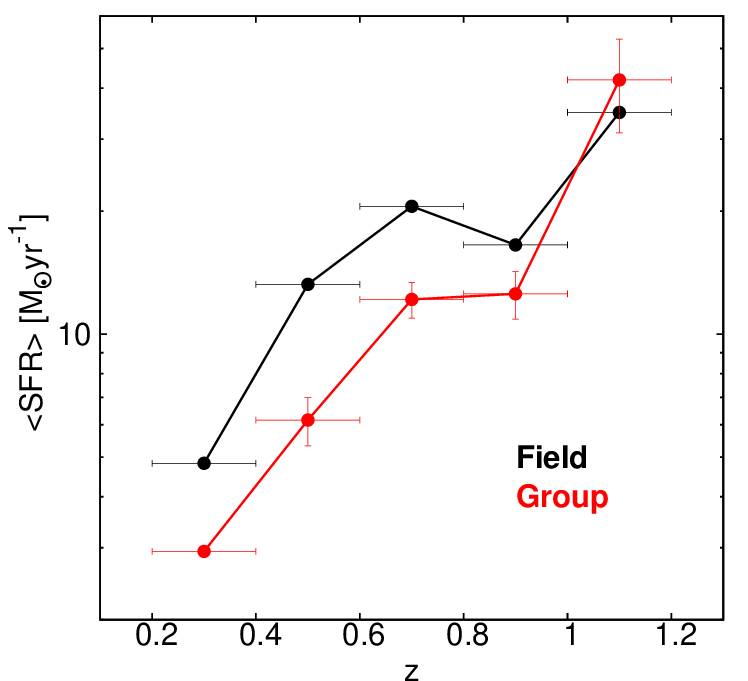}
\caption{ Mean SFR as a function of redshift. Black points show the mean of SFR for galaxies in the whole galaxy population and the red points  and the error bars indicate the mean SFR in bins of redshift and respective errors in the mean for group galaxies.}
\label{meanSFR}
 \end{figure}

By comparing the $\Sigma$SFR$-$M$_{200}$ relation at different redshifts, we see a clear evolution in the level of star formation activity.
 Indeed, the total star formation activity in high redshift groups is higher with respect to the low redshift sample. By dividing the two samples
 in several $M_{200}$ bins, we estimate a mean difference of $0.8\pm0.1$ dex between high and low redshift groups. A milder difference ($0.35\pm0.1$ dex)
 is observed between the [0.15-0.5] redshift bin and the groups at $z < 0.085$ of \cite{Yang2007}. In order to check if this evolution is happening faster
 in the group galaxy population than the field population, we compare the  mean SFR per galaxy in groups  as a function of redshift with the mean SFR per galaxy in the whole galaxy population (Figure \ref{meanSFR}).
 The mean SFR per galaxy in groups is derived by  dividing  the sum of the corrected $\Sigma$SFR for the groups by the sum of their corrected N  in the considered redshift bin.
 For the mean SFR per galaxy in the whole galaxy population, we use the infrared luminosity density obtained by \cite{Gruppioni2013},  based on PACS data, for galaxies with mass above  $M*>10^{10} M_{\odot}$. Using  \cite{Kennicutt1998} relation, we convert the IR luminosity density to the SFR density. In order to obtain the mean SFR per galaxy, we divide the SFR density by the number density for  $M*>10^{10} M_{\odot}$ derived from the integration of the sum of the quiescent and star-forming galaxies mass function derived by \cite{Ilbert2010}. 
 According to Figure \ref{meanSFR}, group galaxies have very similar level of star formation activity with the whole galaxy population at z $\sim$ 1, but at lower redshifts they experience much faster evolution than the global relation.
 Since the whole galaxy population should be dominated by lower mass halos, $M_{200} \sim 10^{12-12.5} M_{\odot}$ according to the predicted dark matter halo mass function (e.g. \citealt{ Jenkins2001, Tinker2008})
 and to the estimate of \cite{Eke2005}, this would imply that the level of SF activity is declining more rapidly since $z\sim 1$ in the more
 massive halos than in the more common lower mass halos consistent with \cite{Ziparo2014}. This confirms a ``halo downsizing'' effect as discussed in \cite{Popesso2012}. In addition, as discussed above, 
 We also point out that the result does not change if we do not calibrate SFR$_{SED}$ (see  \S \ref{sed_fitting}). The effect of this calibration is just to slightly reduce the scatter of the relation.

The central and bottom panels of Figure \ref{SFR} show the N and the $\Sigma$ M$_*$$-$M$_{200}$ relations in the two redshift bins. In these
 cases we see a very tight relation in both samples as confirmed by a Spearman test at 99\% confidence level (see Table \ref{fitMN}). This is not surprising.
 Indeed, while the stellar mass function of the galaxy population, and of the group galaxy population in particular,
 is not evolving significantly since redshift $\sim$1 as shown in \cite{Ilbert2010} and \cite{Giodini2012}, respectively, the SF activity
 of the Universe is dropping down by an order of magnitude in the same time window (see e.g. \citealt{Magnelli2013} for the whole galaxy population,
 \citealt{Popesso2012} for groups and clusters in particular). As a consequence the spread in $\Sigma$SFR is much higher than the spread of $\Sigma$M$_*$.
 Thus, we see a strong correlation between $\Sigma$M$_*$ and M$_{200}$ and only a mild correlation between $\Sigma$SFR and M$_{200}$.

The halo occupation distribution is consistent with a linear relation in the high redshift bin and marginally consistent with it (within 2.5 $\sigma$, see Table \ref{fitMN}) in the low redshift bin. 
This is probably due to a bias induced by our selection of groups with more than 5 members, needed to properly define the group
 redshift and membership. Indeed, this cut makes more likely that we favor the selection of rich groups for a given mass, in particular among the 
low mass groups. Since the mean M$_{200}$ of the low redshift sample is a factor of two lower than the mean mass of the high redshift sample, 
this bias is more significant in the low redshift sample at low masses, leading to a sub-linear halo occupation distribution. Indeed, the halo occupation distribution obtained using the \cite{Yang2007} group
 subsample at $z < 0.085$ and with the same stellar mass cut is highly consistent with a linear relation for
 halos with masses  $M_{200} > 10^{13}M_{\odot}$ as discussed above (see central panel of Figure \ref{yang}). As for the $\Sigma$SFR$-$M$_{200}$ relation,
 also the $\Sigma$ M$_*$$-$M$_{200}$ relation shows a double slope, $\Sigma{M_*} \propto M_{200}^{0.61\pm0.002}$ at $M_{200} < 10^{13}M_{\odot}$
 and $\Sigma{M_*} \propto M_{200}^{1.00\pm0.07}$ at $M_{200} > 10^{13}M_{\odot}$. Since the \cite{Yang2007} groups with masses below $M_{200} < 10^{13}M_{\odot}$ typically contain only the central galaxy,
 the relation below this limit shows actually the mean relation between the central galaxy stellar mass and the halo mass. We should note that different fitting methods on our sample lead to perfectly consistent results.

We point out that, according to \cite{Popesso2007}, groups exhibit a much flatter radial density profile
 with respect to more massive systems. Thus, the correction for projection effects for groups should be higher than for more
 massive systems. However, our sample covers a much lower and narrower mass range with respect to the one of \cite{Popesso2007} and we do
 not know accurately the radial density profile of our group sample. We point out that the correction is of the order
 of 15-20\% and it would not change significantly our results given the relatively large error on the slope of the relation.
 We also notice that the slope of the observed relation is consistent with the one observed in galaxy clusters at much higher
 masses (\citealt{Marinoni2002, Pisani2003, Lin2004, Popesso2007}).

We do not observe any evolution in the halo occupation distribution since $z\sim 1.1$. Similarly we do not observe evolution in the relation between the total stellar mass in
 groups and the total mass, in agreement with the results of \cite{Giodini2012} (see bottom panel of Figure  \ref{SFR} and right panel of Figure \ref{yang}).

The picture emerging from Figure \ref{SFR} and \ref{yang} is that accretion of galaxies or stellar mass goes together with accretion of total halo mass.
Since, the massive halos are not predicted to increase their total halo mass by a large factor ( \citealt{Stewart2008, Fakhouri2010, Moster2013})  through a merger event in the last 10 Gyr, thus, the same is true for their stellar mass
 and number of galaxies. This picture in addition to Figure \ref{meanSFR} imply  that the most evident evolution of the galaxy population of the most massive systems is in terms of the quenching of
 their star formation activity. This also implies that the group galaxy population should progressively move from high to low specific star formation
 rate from $z\sim 1$ to $z\sim 0$ and move away from the Main Sequence more rapidly than galaxies in lower mass halos, in agreement with the result of \cite{Ziparo2014}.

\subsection{Fraction of  MS galaxies vs. M$_{200}$ and velocity dispersion}
\begin{figure}
\includegraphics[width=80mm]{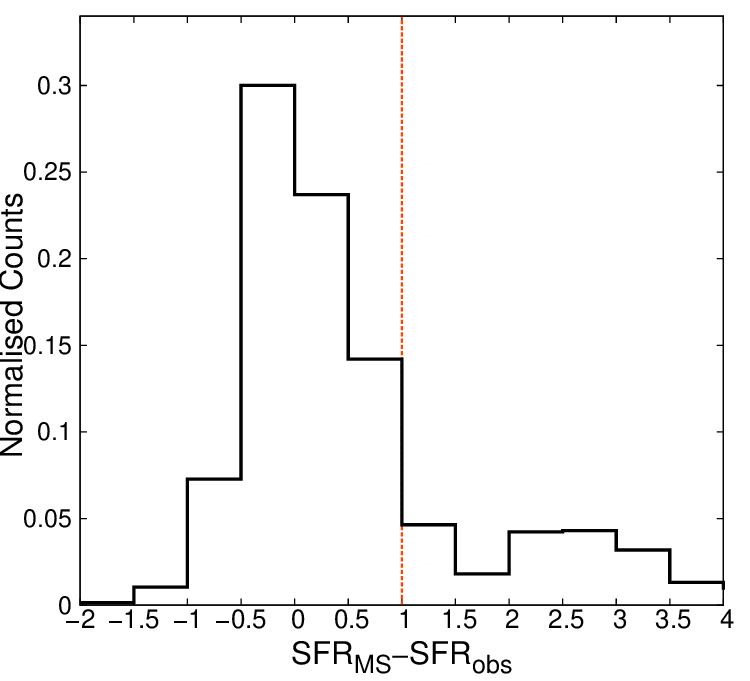}
 \caption{Normalized distribution of differences between Main Sequence SFR and observed SFR  of member galaxies ($\Delta$SFR). The red vertical lines show our limit for separation MS member galaxies.\label{residual}} 
 \end{figure}

\begin{figure}
\includegraphics[width=80mm]{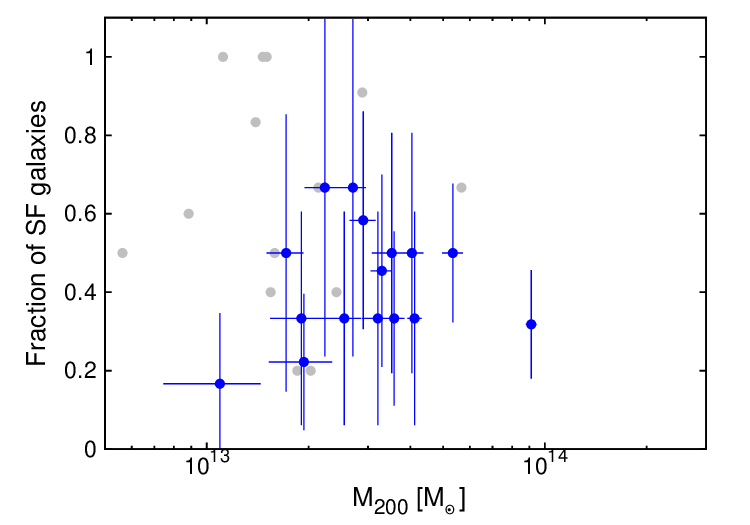}
\caption{Fraction of star-forming galaxies as a function of halo mass for the low-z sample with more
 than 10 members (blue points) and less than 10 members (in grey). Spearman test confirms no correlation for this sample. \label{MxFl}}
 \end{figure}

\begin{figure}
\includegraphics[width=80mm]{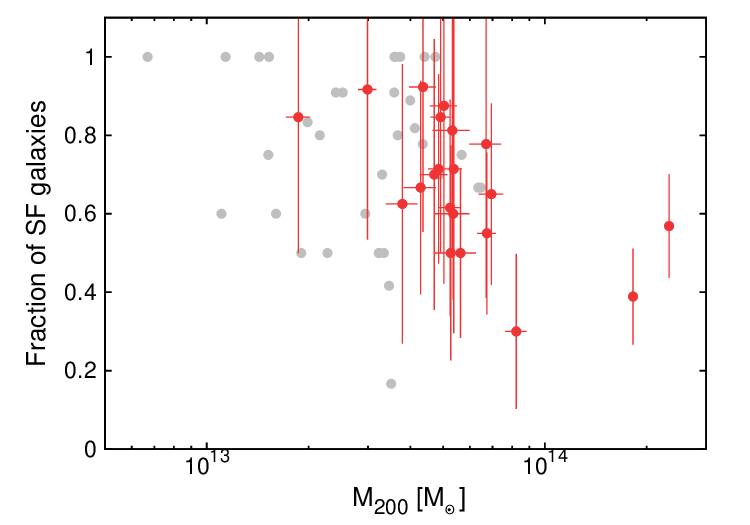}
 \caption{Fraction of star-forming galaxies as a function of halo mass for the high-z sample with more
 than 10 members (red points) and less than 10 members (in grey). Spearman test confirms a significant anticorrelation for the this sample. \label{MxF}}
 \end{figure}

\begin{figure}
\includegraphics[width=80mm]{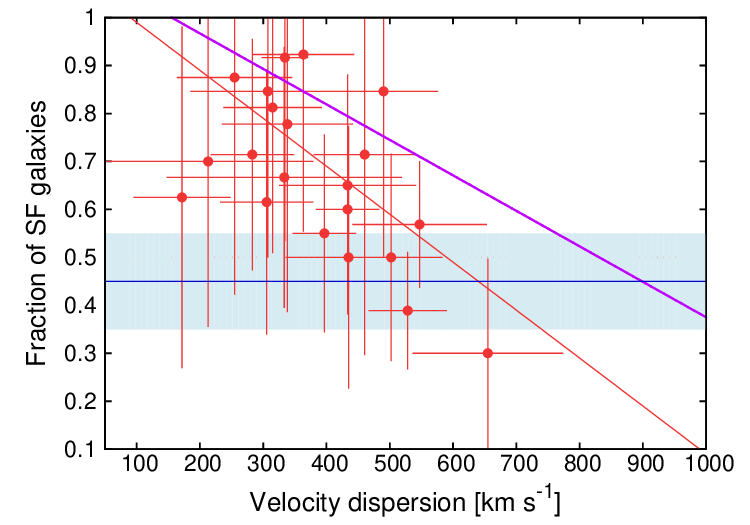}

\caption{Fraction of star-forming galaxies vs. velocity dispersion for groups in the high-z sample with more than 10 spectroscopic members. The magenta line is the upper envelope of Poggianti et al. (2006)
for the EDisCS clusters and groups at z=0.4-0.8. The horizontal blue line and the shaded blue area show the median fraction of star-forming galaxies and its corresponding one sigma error in low-z groups. \label{vF}}
 \end{figure}

Often the level of star formation activity in groups and clusters is estimated through the fraction of star-forming galaxies.
 In order to compare with previous results, we analyse in this section the evolution of the fraction of star-forming galaxies as a function
 of the group halo mass. We define the star-forming galaxies as the ones lying
 on the Main Sequence (\citealt{Elbaz2007}). In order to identify the main sequence at different
 redshifts, we extrapolate the MS relation at the mean redshift of each redshift bin by interpolating
 the MS relation of \cite{Peng2010}, \cite{Noeske2007a} and \cite{Elbaz2007}. According to these
 works the scatter of the relation is $\sim$ 0.3 dex.  Figure \ref{residual} illustrates the distribution of 
the residual $\Delta$(SFR)=SFR$_{MS}$$-$SFR$_{observed}$, where SFR$_{MS}$ is the SFR given by the MS relation at a given mass and SFR$_{observed}$ is the observed SFR of a galaxy at that mass. The distribution shows a well known
bimodal distribution with the Gaussian representing the MS location with peak around
0 residual, and a tail of quiescent/low star-forming galaxies at high positive values of
$\Delta$SFR. This distribution is reminiscent of the bimodal behavior of the U-R galaxy color
distribution observed by \cite{Strateva2001} in the SDSS galaxy sample. 
At all redshifts, the value $\Delta$SFR = 1 turns out to be the best separation for MS galaxies. It is also consistent with 3$\sigma$ of main sequence uncertainty.
The fraction of star-forming galaxies is, then, defined as the ratio between the number of SF galaxies with M$_*>10^{10} M_{\odot}$ and the total number of galaxies
 with M$_*>10^{10} M_{\odot}$. We apply the same spectroscopic incompleteness correction for the number of star-forming galaxies as for the total number of galaxies , so it is cancelled from the fraction. We do not find any correlation in the low redshift bin with the halo mass (see Table \ref{fitMN} and Figure \ref{MxFl} ). This
 is confirmed also by a lack of correlation in the \cite{Yang2007} group subsample at $z < 0.085$.
 We observe a significant anti-correlation with the halo mass in the high redshift bin, as confirmed by
 a Spearman test (see Figure \ref{MxF}). Figure \ref{vF} shows the relation between fraction of star-forming galaxies and velocity dispersion
 for the galaxy groups with more than 10 spectroscopic members for which we have a reasonable estimate of the galaxy velocity dispersion. The magenta line in Figure \ref{vF}, is the upper envelope of \cite{Poggianti2006} for
 the EDisCS clusters and groups at z=0.4-0.8. Even in this case,
 high mass systems seem to be already evolved at z$\sim$1 by showing a fraction of star-forming
 galaxies consistent with the low redshift counterparts at $z < 0.085$, where we measure a mean constant fraction of SF galaxies of $0.28\pm 0.5$.

 Given the almost linear relation between the $\Sigma$SFR and M$_{200}$ in the high-z sample, this implies that most of the contribution to the total
 SFR of the most massive systems ($M_{200} \sim 10^{14}M_{\odot}$) is given by few but highly star-forming galaxies,
 while in lower mass systems ($M_{200} \sim 10^{13}M_{\odot}$) it is given by more star-forming galaxies of average activity. Thus, this would still indicate a faster evolution in the more massive
 systems in terms of star formation activity with respect to lower mass groups.


\subsection{ Comparison with the mock catalog}

\begin{figure}
\includegraphics[width=80mm]{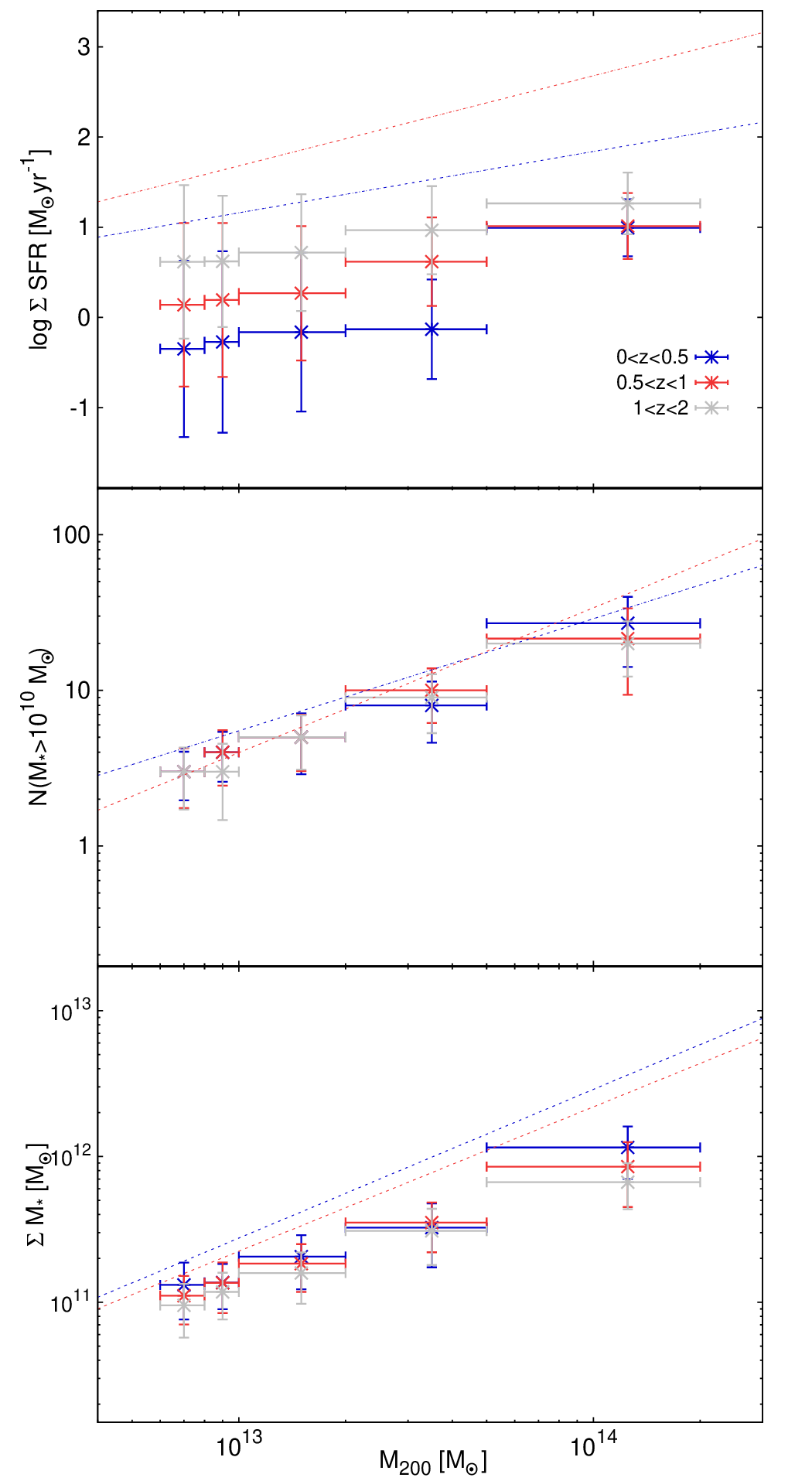}
 \caption{$\Sigma$ SFR- (upper panel), N- (middle panel) and $\Sigma$ M$_*$- (bottom panel)  M$_{200}$ relations for the groups with
 0$<$z$<$0.5 (in red) and 0.5$<$z$<$1 (in blue) and with 1$<$z$<$2 (in grey) for the mock catalog. The dashed lines show the results based on the observations.\label{Mockresults}}

 \end{figure}

\begin{figure*}
\includegraphics[width=150mm]{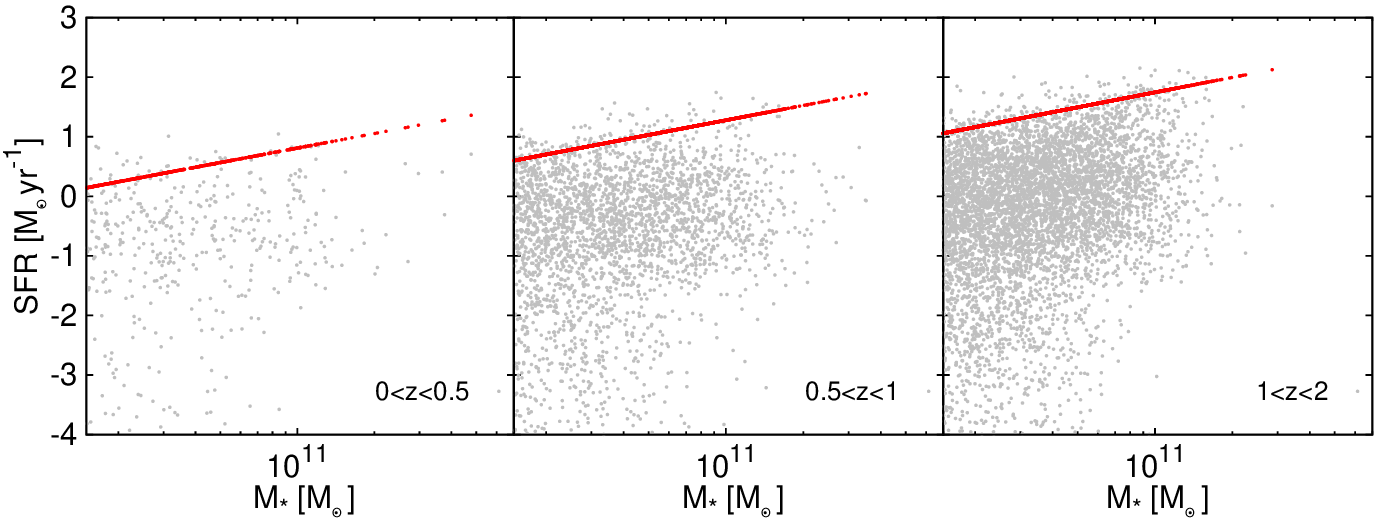}
\caption{SFR as a function of stellar mass for the member galaxies in the mock catalog. The red points show the position of the main sequence for the lowest redshift (z= 0, 0.5 and 1 from left to right, respectively) in each bin.\label{mainsequence}}

 \end{figure*}

To compare our results with theory, we look at the results based on the mock catalog of the Millennium Simulation as described
 in Sect. \ref{mock}. We analyse the same relations studied in our work by extracting from the mock catalog a sample of groups
 in the same mass range and redshift range adopted in our study. The quantities $\Sigma$ M$_*$, $\Sigma$ SFR and number of
 galaxies per halo mass are calculated by following the same criteria used for the real dataset. In addition, we also estimate
 the properties of the groups at 1$<$z$<$2 to completely follow the evolutionary trends of galaxies up to z $\sim$ 2. Figure
 \ref{Mockresults} shows the predictions of the same relations presented in Figure \ref{SFR}.  The top panel of the figure shows
 the total SFR of the mock groups as a function of their halo masses. As already known, the semi-analytical models of the Millennium
 simulation underpredict the level of star formation activity of the galaxy population and, in particular, of the group and cluster
 galaxies. Indeed, even the level of activity of the high redshift groups is well below the level of the low redshift groups of our
 sample (dotted blue line in the plot). This class of models assumes that, when galaxies are accreted onto a more massive system,
 the associated hot gas reservoir is stripped instantaneously. This, in addition to the AGN feedback, induces a very rapid decline
 of the star formation histories of satellite galaxies, and contributes to create an excess of red and passive galaxies with respect
 to the observations (e.g. \citealt{Wang2007}). More recent high resolution simulations do not help in improving the results
 (\citealt{Weinmann2011, Guo2011}). This is known as the ``overquenching problem" for satellites galaxies. Over 95\%
 of the cluster and group galaxies within the virial radius in the local simulated Universe are passive (\citealt{Guo2011}),
 at odds with observations (e.g. \citealt{ Weinmann2006, Kimm2009, Liu2010, Hansen2009, Popesso2005}).
 Indeed, as Figure \ref{mainsequence} shows, galaxies in mock groups reside under the main sequence in any redshift bin, indicating that
 the evolution even in group galaxies is happening at $z > 2$. This is at odds with our results since in the previous section we have shown
 that in the low mass groups most of the galaxies above $10^{10}M_{\odot}$ are Main Sequence galaxies.

We do not observe any evolution in the halo occupation distribution (central panel of Figure \ref{Mockresults}), which is consistent also quantitatively with
 the halo occupation distribution observed in our group sample. In the same way we do not observe any evolution in the $\Sigma$ M$_*-M_{200}$ relation but
 we also observe a quantitative discrepancy with respect to observations. Indeed, at any redshift the total stellar mass in groups is
 underpredicted with respect to the observed one. This is understandable given the much lower star formation rate of the simulated
 group galaxies with respect to the observations, which limits the galaxy stellar mass growth.





\section{Summary and conclusion}

In this paper we provide an analysis of the evolution of the total star formation activity, total stellar
 mass and halo occupation distribution by using one of the largest X-ray selected samples of galaxy groups with secure spectroscopic
 identification on the major deep field surveys (ECDF, CDFN, COSMOS, AEGIS) up to z$\sim$1.1. We first check the robustness of our method in determining the group velocity dispersion and
 membership extensively using mock catalogs and check the possible biases induced by the spectroscopic incompleteness of the surveys used
 in our analysis. We show that for a robust measurement of the group velocity dispersion and group membership
 definition a first guess of the velocity dispersion derived from the X-ray luminosity is essential
 for a reliable result. We compare our results with the one based on an optically-selected sample of groups at
 $z< 0.085$ in order to fully follow the evolution of the galaxy population in groups to the local Universe.
We list below our main results:
\begin{itemize}
\item [-] We observe a clear evolution in the level of star formation activity in galaxy groups. Indeed, the total star formation
 activity in high redshift groups (0.5 $<$ z $< $1.1) is higher with respect to the low redshift sample (0.15 $<$ z $< $0.5) at any mass by almost $0.8 \pm 0.1$ dex.
 A milder difference ($0.35 \pm 0.1$ dex) is observed between the [0.15-0.5] redshift bin and the groups at $z < 0.085$. This
 evolution seems to be much faster than the one observed in the whole galaxy population (\citealt{Gruppioni2013}), dominated by lower mass halos
 (M$_{200} \sim 10^{12-12.5}$ M$_{\odot}$,  \citealt{Jenkins2001, Tinker2008, Eke2005}).
 This would imply that the level of SF activity is declining more rapidly since $z\sim 1.1$ in the more massive halos than in
 the more common lower mass halos, confirming a ``halo downsizing'' effect as discussed by \cite{Popesso2012}.
\item[-] The halo occupation distribution and the total stellar mass$-$M$_{200}$ relation are consistent with a linear relation in all redshift bins in the
 M$_{200}$ range considered in our analysis. We do not observe any evolution in the halo occupation distribution since $z \sim 1$. Similarly we do not
 observe evolution in the relation between the total stellar mass in groups and the total mass, in agreement with the results
 of \cite{Giodini2012}. The picture emerging from our findings is that massive groups at M$_{200} \sim 10^{13-14}$M$_{\odot}$
 have already accreted the same amount of mass and have the same number of galaxies as the low redshift counterparts, as predicted
 by  \cite{Stewart2008}. This implies that the most evident evolution of the galaxy population of the most massive systems acts
 in terms of quenching their galaxy star formation activity. This also implies that the group galaxy population should progressively
 move from high to low specific star formation rates from $z \sim 1$ to $z \sim 0$ and rapidly move away from the Main Sequence since
 $z \sim 1$ consistent with the recent results of \cite{Ziparo2013} based on a similar dataset.
\item[-] The analysis of the evolution of the fraction of SF galaxies as a function of halo mass or velocity dispersion shows that
 high mass systems seem to be already evolved at z$\sim$1 by showing a fraction of star-forming galaxies consistent with the low
 redshift counterparts at $z < 0.085$.  Given the almost linear relation between the $\Sigma$SFR  and M$_{200}$ in the high-z sample, this implies that most
 of the contribution to the total SFR of the most massive systems ($M_{200} \sim 10^{14}M_{\odot}$) is given by few highly star-forming galaxies, while in lower mass systems ($M_{200} \sim 10^{13}M_{\odot}$) is given by many galaxies of average activity.
 This would be an additional sign of a faster evolution in the more massive systems in terms of star formation
 activity with respect to lower mass groups. Thus, it would confirm the ``halo downsizing'' effect.
\item[-] The comparison of our results with the prediction of the Millennium Simulation semi-analytical model confirms the known
 problem of the models. We confirm the strong bias due to the ``satellite overquenching'' problem in suppressing significantly
 the SF activity of group galaxies (more than an order of magnitude) at any redshift with respect to observations. The halo occupation distribution
 predicted by the simulations is remarkably in agreement with the observations. But due to the low SF activity of galaxies
 in massive halos, the models predict also a lower total stellar mass in groups with respect to the observed one at any redshift.
\end{itemize}

Our results support a scenario in which the quenching of SF occurs earlier in galaxies embedded
 in more massive halos, though we are considering a quite narrow halo mass range. This would be consistent
 with the results obtained by \cite{Popesso2012} in a similar redshift range but in a broader mass range, which
 includes also galaxy clusters. Other evidences in the literature support the differential evolution of the
 SF activity in massive halos with respect to the field or lower mass halos. For instance, the formation of the
 galaxy red sequence, which leads to the local dichotomy between red and blue galaxies, happens earlier in
 groups than in the field especially at high stellar masses (\citealt{Iovino2010, Kovach2010, Mok2013, Wilman2009, Wilman2012}).
 Morphological transformations are in place in groups at z$<$1, leading to a transient population of ``red spirals'' not observed
 in the field (\citealt{Balogh2009, Wolf2009, Mei2012}). There is also evidence that at z$\sim$1 there
 is a flattening of the SFR-density relation (\citealt{Elbaz2007, Popesso2011, Cooper2008, Ziparo2014})
 with respect to the local anti-correlation. \cite{Ziparo2014} find on the very same dataset that the differential
 evolution of the groups galaxies with respect to field is due to the fact that star-forming group galaxies are perfectly
 on the Main Sequence at z$\sim$1 whereas at lower redshift they are quenched, thus, dropping off the MS quicker than field
 galaxies towards the region of SF quiescence.

What is causing this differential evolution as a function of the halo mass? According to \cite{Peng2010} massive galaxies,
 as the ones considered in our sample, evolve mostly because of an internally driven process, called 'mass quenching', caused
 perhaps by feedback from active galactic nuclei. But since this process is unlikely to be more efficient in quenching SF of
 massive galaxies in massive halos than in other environments as the stellar mass functions do not change significantly in groups with respect to field (\citealt{Giodini2012}),
 the ``environmental quenching'' must be the main mechanism for
 quenching the SF of the most massive satellites in massive halos. Which kind of process is causing this ``environmental quenching''
 is still quite unknown. Ram-pressure stripping (\citealt{Gunn1972}) and starvation \citep{Larson1980} are two plausible candidates for
 producing this quenching.  Ram-pressure stripping  ``quench"  star formation immediately \citep{Abadi1999} as it  can sweep Interstellar medium 
out of a galaxy. Starvation, caused by the removal of the hot gas halo reservoirs of galaxies which leads to a cut in the supply of cold gas in
 the galaxy is also a likely candidate. Tidal galaxy-galaxy encounters or  the interaction with the intra-cluster/intra-group medium can lead to
 the removal of galaxy hot gas reservoirs which induce starvation. Therefore, starvation should quench SF earlier in more massive halos than in low mass halos, as we observe.

 \cite{Cen2011} propose that this differential evolution could be explained simply in terms of the current theory of gas
 accretion that hinges on the cold and hot two-mode accretion model (\citealt{Keres2005, Dekel2006}). The halo mass
 is the main determinant of gas accretion: large halos primarily accrete hot gas while small halos primarily accrete cold gas.
 The overall heating of cosmic gas due to formation of large halos (such as groups and clusters) and large-scale structure
 causes a progressively larger fraction of halos to inhabit regions where gas has too high entropy to cool to continue feeding
 the residing galaxies. The combined effect is differential in that overdense regions are heated earlier and to higher
 temperatures than lower density regions at any given time. Because larger halos tend to reside in more overdense regions
 than smaller halos, the net differential effects would naturally lead to both the standard galaxy downsizing effect and the halo downsizing effect.

The current analysis can not provide evidences in favour of one of these scenarios. Further analysis must be conducted to study the
 cold gas content of galaxies in halos of different masses, to distinguish between the different possibilities and identify the process responsible for the ``environmental quenching''.

\appendix
\section{X-ray groups of galaxies in CDFN}
The catalog of X-ray groups follows the original results of \cite{Bauer2002}, based on the first 1Ms Chandra data. The main difference in the catalog consist in a self-consistent use
 of the flux at R$_{500}$, larger apertures for the flux extraction. This allows us to use our calibrations of group masses, provided by COSMOS (\citealt{Leauthaud2010}) and ECDFS (Finoguenov et al. subm.) surveys.
In column 1, 2 and 3, we provide the group identification number, RA and Dec. of the peak of X-ray emission. In Column 4, the mean of red sequence redshifts which is substituted with the median of spectroscopic redshift in
 case there is a spectroscopic redshift determination for the group member galaxies is given. The group flux in the 0.5--2 keV band in Column 5 with the corresponding
 1$\sigma$ error is listed. The rest-frame luminosity in the 0.1--2.4 keV is presented in Column 6. Column 7 gives the estimated total mass, M$_{200}$, computed following  \cite{Leauthaud2010}
 and assuming a standard evolution of scaling relations: M$_{200}E_z=f(L_xE^{-1}_z)$ where $E_z=(\Omega_M(1+z)^3+\Omega_\Lambda)^{1/2}$,
 standard evolution of the scaling relation. The corresponding r$_{200}$ in degrees is listed in Column 8. Column 9 lists flux 
 significance which provides insights on the reliability of both the source detection and the identification. Column 10
 presents the flag for our identification, as described in section \ref{Xray}. The velocity dispersion estimated from X-ray luminosities is given in column 11.
 The number of spectroscopic member galaxies inside 2$\times$ r$_{200}$ is given in Column 12.

\begin{table*}
\begin{center}
\caption{ X-ray group catalog:(1) X-ray ID; (2) RA [deg]; (3) Dec[deg]; (4) z; (5) Flux [$10^{-15}ergcm^{-2}s^{-1}$]; (6)$L_X(0.1-2.4 keV)[10^{42}erg/s]$; (7) $M_{200}$[$10^{13}M_\odot$];
 (8) $r_{200}$[deg]; (9) Flag; (10) Flux significance; (11) Velocity dispersion from X-ray luminosities [km/s]; (12) N(z)}
\begin{tabular}{lcccccccccccc}
 \hline\hline\\[1ex]
ID & RA & Dec & z & Flux & $L_X$ & $M_{200}$ & $r_{200}$ & Flux & Flag & Velocity  & N($z_{spec}$) \\
   &    &     &   &      &       &           &           &  significance  &     & dispersion       \\
(1)&(2)&(3)&(4)&(5)&(6)&(7)&(8)&(9)&(10)&(11)&(12)\\[1ex]
 \hline\\[1ex]
2 & 189.45619 & 62.36314 & 0.398 & 1.26$\pm$0.36 & 1.21$\pm$0.34 & 2.16$\pm$0.35 & 0.0262 & 3.52 & 3 & 229 & 0\\
4 & 189.26089 & 62.35124 & 0.800 & 0.54$\pm$0.17 & 3.60$\pm$1.12 & 2.94$\pm$0.52 & 0.0177 & 3.19 & 1 & 277 & 9 \\
5 & 188.86385 & 62.35366 & 0.652 & 1.91$\pm$0.57 & 6.32$\pm$1.89 & 4.69$\pm$0.79 & 0.0238 & 3.34 & 3 & 319 & 0\\
6 & 189.36276 & 62.32381 & 0.277 & 0.84$\pm$0.19 & 0.34$\pm$0.08 & 1.12$\pm$0.15 & 0.028 & 4.22 & 2 & 176 & 9  \\
7 & 189.48284 & 62.25552 & 0.455 & 2.99$\pm$0.24 & 3.83$\pm$0.31 & 4.12$\pm$0.19 & 0.0294 & 12.22 & 1 & 293 & 14  \\
8 & 189.18499 & 62.26416 & 0.850 & 1.38$\pm$0.17 & 8.92$\pm$1.07 & 4.85$\pm$0.34 & 0.0202 & 8.36 & 1 & 336 & 46 \\
9 & 188.98803 & 62.2646 & 0.375 & 1.07$\pm$0.28 & 0.88$\pm$0.23 & 1.82$\pm$0.27 & 0.0259 & 3.78 & 3 & 214 & 3 \\
10 & 189.07392 & 62.26007 & 1.999 & 0.39$\pm$0.85 & 36.09$\pm$7.74 & 4.41$\pm$0.54 & 0.0119 & 4.66 & 2 & 402 & 3 \\
13 & 189.0872 & 62.18605 & 1.014 & 0.51$\pm$0.18 & 7.12$\pm$2.58 & 3.67$\pm$0.75 & 0.0164 & 2.76 & 1 & 314 & 20\\
14 & 189.5959 & 62.1628 & 0.914 & 1.24$\pm$0.32 & 10.77E$\pm$2.74 & 5.13$\pm$0.75 & 0.0196 & 3.92 & 3 & 348 & 0 \\
15 & 189.33336 & 62.12823 & 0.943 & 0.76$\pm$0.17 & 7.80$\pm$1.71 & 4.13$\pm$0.52 & 0.0179 & 4.56 & 3 & 323 & 0\\
16 & 189.13775 & 62.15006 & 0.840 & 0.48$\pm$0.12 & 3.77$\pm$0.93 & 2.92$\pm$0.41 & 0.0171 & 4.05 & 1 & 279 & 12 \\
17 & 189.04209 & 62.14711 & 1.139 & 0.61$\pm$0.14 & 11.41$\pm$2.52 & 4.37$\pm$0.56 & 0.0162 & 4.52 & 3 & 343 & 2 \\
19 & 188.96164 & 62.12097 & 0.491 & 2.49$\pm$0.40 & 3.84$\pm$0.62 & 4.00$\pm$0.38 & 0.0275 & 6.2 & 3 & 292 & 0 \\
20 & 189.538 & 62.13181 & 0.948 & 0.64$\pm$0.23 & 6.89$\pm$2.46 & 3.81$\pm$0.77 & 0.0174 & 2.8 & 3 & 314 & 0 \\
21 & 188.86226 & 62.10217 & 0.895 & 1.37$\pm$0.41 & 11.05$\pm$3.41 & 5.30$\pm$0.93 & 0.0201 & 3.24 & 5 & 351 & 0 \\
22 & 189.11361 & 62.10088 & 1.217 & 0.45$\pm$0.13 & 10.99$\pm$3.14 & 4.00$\pm$0.65 & 0.0152 & 3.5 & 3 & 337 & 1 \\
23 & 189.28445 & 62.09072 & 0.956 & 0.69$\pm$0.17 & 7.46$\pm$1.88 & 3.97$\pm$0.57 & 0.0175 & 3.96 & 3 & 319 & 0 \\
24 & 189.02017 & 62.08888 & 1.217 & 0.91$\pm$0.22 & 17.94$\pm$4.34 & 5.37$\pm$0.74 & 0.0167 & 4.13 & 5 & 375 & 0 \\
25 & 189.22003 & 62.07086 & 0.188 & 4.20$\pm$0.53 & 0.65$\pm$0.08 & 1.75$\pm$0.13 & 0.045 & 7.92 & 3 & 204 & 0 \\
27 & 189.28874 & 62.02523 & 1.640 & 1.20$\pm$0.27 & 46.55$\pm$10.55 & 6.73$\pm$0.88 & 0.0152 & 4.41 & 5 & 442 & 0 \\
28 & 189.17982 & 62.02048 & 0.426 & 1.80$\pm$0.42 & 1.98$\pm$0.47 & 2.84$\pm$0.39 & 0.0273 & 4.22 & 3 & 254 & 0 \\
30 & 189.08941 & 62.26975 & 0.681 & 0.17$\pm$0.11 & 0.90$\pm$0.61 & 1.42$\pm$0.52 & 0.0155 & 1.48 & 2 & 207 & 7 \\
31 & 189.10007 & 62.25822 & 0.642 & 0.45$\pm$0.28 & 1.71$\pm$1.06 & 2.16$\pm$0.73 & 0.0185 & 1.6 & 2 & 240 & 10 \\
32 & 189.09046 & 62.26367 & 1.241 & 0.19$\pm$0.08 & 6.75$\pm$2.73 & 2.93$\pm$0.66 & 0.0135 & 2.48 & 4 & 302 & 3\\
33 & 189.3379 & 62.15165 & 1.126 & 0.49$\pm$0.11 & 9.45$\pm$2.23 & 3.95$\pm$0.53 & 0.0158 & 4.24 & 3 & 330 & 0 \\
34 & 189.53013 & 62.11978 & 0.280 & 1.26$\pm$0.57 & 0.51$\pm$0.23 & 1.41$\pm$0.36 & 0.03 & 2.2 & 5 & 192 & 0 \\
\hline
\label{biastable100} 
\end{tabular}
\end{center}
\end{table*}

\section*{Acknowledgements}
GE acknowledges the support from and participation in the International Max-Planck Research School on Astrophysics at the Ludwig-Maximilians University.
MLB acknowledges financial support from an NSERC Discovery Grant, and from NWO and NOVA grants that supported his sabbatical at the University of Leiden, where this work was completed.
PACS has been developed by a consortium of institutes led by MPE 
(Germany) and including UVIE (Austria); KUL, CSL, IMEC (Belgium); CEA, OAMP (France); MPIA (Germany); IFSI, OAP/AOT, OAA/CAISMI, LENS, SISSA 
(Italy); IAC (Spain). This development has been supported by the funding 
agencies BMVIT (Austria), ESA-PRODEX (Belgium), CEA/CNES (France),
DLR (Germany), ASI (Italy), and CICYT/MCYT (Spain).

This research has made use of NASA's Astrophysics Data System, of NED,
which is operated by JPL/Caltech, under contract with NASA, and of
SDSS, which has been funded by the Sloan Foundation, NSF, the US
Department of Energy, NASA, the Japanese Monbukagakusho, the Max
Planck Society, and the Higher Education Funding Council of England.
The SDSS is managed by the participating institutions
(www.sdss.org/collaboration/credits.html).

We gratefully acknowledge the contributions of the entire
COSMOS collaboration consisting of more than 100 scientists.
More information about the COSMOS survey is available at
http://www.astro.caltech.edu/∼cosmos.
This project has been supported by the DLR grant 50OR1013 to MPE.\\
\\
\\

\end{document}